\documentclass[traditabstract]{aa}
\usepackage{amsmath}
\usepackage{txfonts}
\usepackage{graphicx}
\usepackage{natbib}
\usepackage{amsfonts}
\usepackage{amssymb}
\usepackage{hyperref}
\usepackage{xcolor}
\usepackage{longtable}
\usepackage{appendix}

\bibpunct{(}{)}{;}{a}{}{,} 

\renewcommand{\hat}[1]{\oldhat{\mathbf{#1}}}

\newcommand{\numax}{\nu_{\mathrm{max}}}
\newcommand{\nucut}{\nu_{\mathrm{cut}}}
\newcommand{\omegacut}{\omega_{\mathrm{cut}}}
\newcommand{\teff}{T_{\mathrm{eff}}}
\newcommand{\dnumodes}{\Delta \nu_{\mathrm{modes}}}

\begin{document}
\title{Analysis of the acoustic cut-off frequency and HIPs in six \emph{Kepler} stars with stochastically excited 
pulsations}


\author{A.~Jim\'enez \inst{1, 2} \and
R.~A. Garc\'\i a\inst{3} \and 
F. P\'erez Hern\'andez \inst{1,2} \and
S.~Mathur\inst{4}
}

\institute{Ins\-ti\-tu\-to de Astrof\'\i sica de Canarias, 38205, La Laguna, Tenerife, Spain
\and Uni\-ver\-si\-dad de La Laguna, Dpto. de Astrof\'isica, 38206, Tenerife, Spain
\and Laboratoire AIM, CEA/DSM -- CNRS - Univ. Paris Diderot -- IRFU/SAp, Centre de Saclay, 91191 Gif-sur-Yvette Cedex, France
\and Space Science Institute, 4750 Walnut Street, Suite 205, Boulder, CO 80301, USA
}

\date{Received  / Accepted }
\abstract{Gravito-acoustic modes in the Sun and other stars propagate in resonant cavities with a frequency below 
a given limit known as the cut-off frequency. At higher frequencies, waves are no longer trapped in the stellar interior and 
become traveller waves. 
In this article we study six 
pulsating solar-like stars at different evolutionary stages observed by the NASA \emph{Kepler} mission. These high 
signal-to-noise targets show a peak structure that extends at very high frequencies and are good candidates for studying the 
transition region between the modes and the interference peaks or pseudo-modes. Following the same methodology successfully 
applied on Sun-as-a-star measurements, we uncover the existence of pseudo-modes in these stars with one or 
two dominant interference patterns depending on the evolutionary stage of the star. We also infer their cut-off frequency as 
the midpoint between the last eigenmode and the first peak of the interference patterns. 
By using ray theory we show that, while the period of one of the interference pattern is very close to half the large 
separation the other, one depends on the time phase of mixed waves, thus carrying additional information on the stellar 
structure and evolution.}
\authorrunning{Jim\'enez, Garc\'\i a, P\'erez Hern\'andez \& Mathur}
\titlerunning{Acoustic cut-off frequency and  HIPs with \emph{Kepler}}
\maketitle

\section{Introduction}
\label{sect:intro}
Solar-like oscillation spectra are usually dominated by p-mode eigenfrequencies corresponding to waves trapped in 
the stellar interior with frequencies below a given cut-off frequency, $\nucut$. However, the oscillation power spectrum of 
the resolved Sun shows regular peak structure that extends well above $\nucut$ \citep[e.g.][]{1988ESASP.286..279J,1988ApJ...334..510L,1991ApJ...373..308D}.
This signal is interpreted as travelling waves whose interferences produce a well-defined pattern corresponding to the 
so-called pseudo-mode spectrum \citep{1990LNP...367...87K}. The amount and quality of the space data provided by the SoHO 
satellite \citep{DomFle1995} allowed us to measure these High-Frequency Peaks (HIPs) using Sun-as-a-star observations 
\citep{GarPal1998} from GOLF \citep{GabGre1995} and VIRGO \citep{1995SoPh..162..101F} instruments. 
The change in the frequency pattern between the acoustic and the pseudo-modes enabled the proper determination of the solar cut-off 
frequency \citep[][]{2006ApJ...646.1398J}.

Theoretically, the cut-off frequency approximately scales as $g \sqrt{\mu/\teff}$, where $\teff$ is the effective temperature, $g$ is the gravity, and $\mu$ the mean molecular weight, with all values measured at the surface. Hence, the observed $\nucut$ can be used to constrain the fundamental stellar parameters.
However, solar observations showed that $\nucut$ changes, for example, with the solar magnetic activity cycle \citep{2011ApJ...743...99J}. Therefore, the accuracy of the scaling relation given above needs to be discussed not only in a theoretical context but also taking into account  observational constraints.

For stars other than the Sun, the detection of HIPs is challenging because of the shorter length of available datasets and the 
lower signal-to-noise ratio (SNR).  
Nevertheless, in this study we report on the analysis of the high-frequency part of the spectrum of six 
pulsating stars observed by \emph{Kepler} \citep{2010Sci...327..977B} for which we were able to characterize the HIP pattern and the cut-off frequency. The time series analysis and the preparation of the spectra are detailed in Sect.~2. In Sect.~3 we describe how to estimate $\nucut$ and compare it with our theoretical expectations. In Sect.~4 we perform a detailed analysis of the HIPs and interpret the results as a function of the evolutionary stage of the stars. Finally we provide our conclusions in Sect.~5.

\section{Data Analysis}
In this study ultra high-precision photometry obtained by NASA's \emph{Kepler} mission has been used to study the high-frequency region of six stars with solar-like pulsations. The stellar identifiers from the \emph{Kepler} Input Catalogue names \citep[KIC,][]{2011AJ....142..112B} are given in the first column of Table~1. Short-cadence time series \citep[sampling rate of  58.85s,][]{2010ApJ...713L.160G} up to quarter 17 have been corrected for instrumental perturbations and properly stitched together using the \emph{Kepler} Asteroseismic Data Analysis and Calibration Software \citep[KADACS,][]{2011MNRAS.414L...6G}. 

\begin{figure}[!htb]
\begin{center}
\includegraphics[scale=0.364,angle=0]{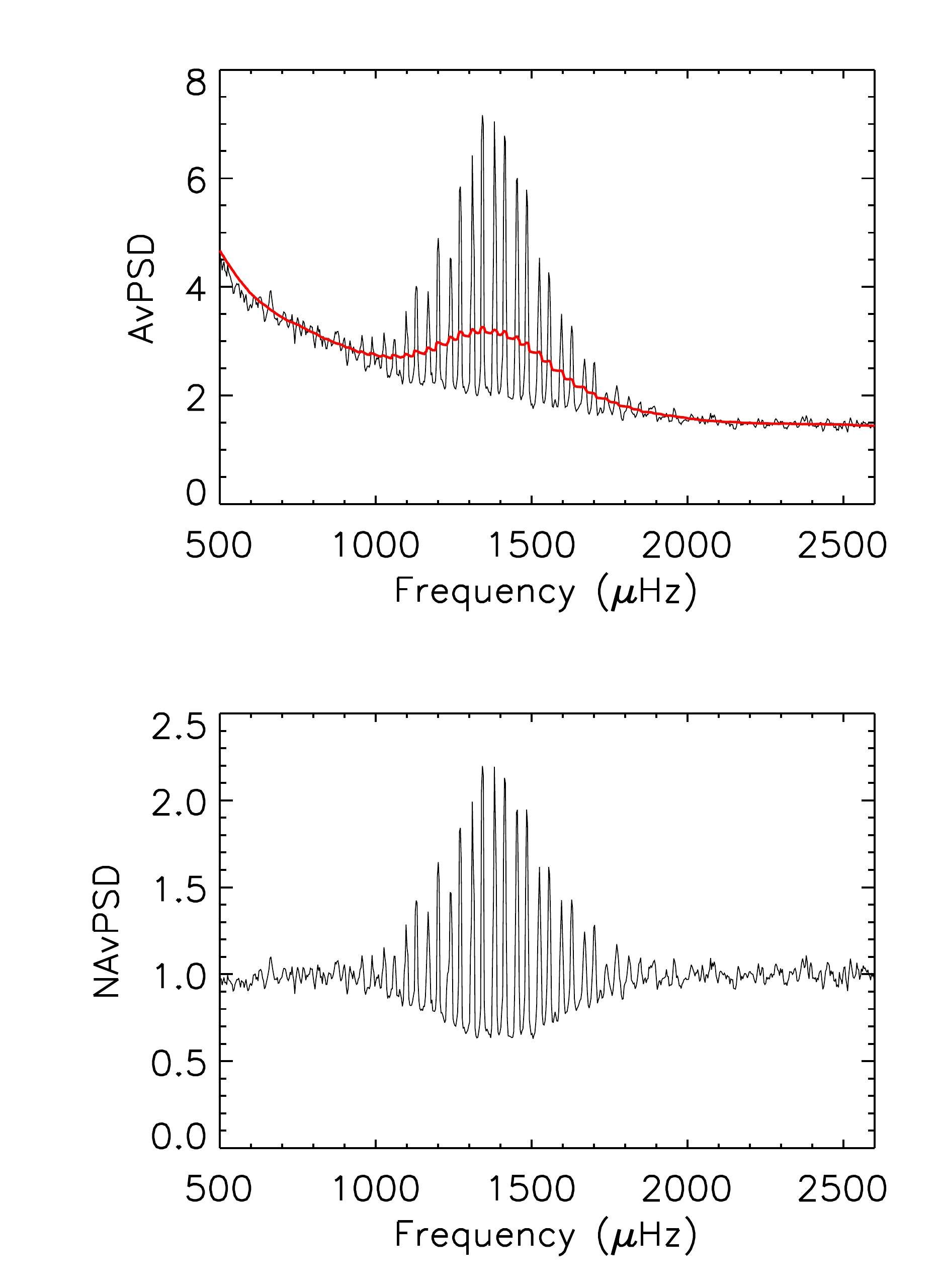}
\caption{Top panel: Smoothed averaged power spectral density of 4-day subseries of KIC~11244118. The red line is the severe smoothing used to normalize the spectrum (NAvPSD) shown in the lower panel (see text for details).}
\label{fig:AvPSD}
\end{center}
\end{figure}

For each star the long time series ($\sim$4 years) are divided into consecutive subseries and the average of all the power 
spectral density (AvPSD) is computed in order to reduce the high-frequency noise in the spectrum. As in the solar case 
(e.g. \citealt[]{GarPal1998,2005ApJ...623.1215J}, \citeyear{2006ApJ...646.1398J}) subseries of $\sim$4 days ($4 \times 1440$ points, 3.92 days)
were used because they are a good compromise between frequency resolution (which improves with longer subseries) and the 
increase of the SNR (which improves with the number of averaged spectra). For the same reasons (as also done in the solar 
case), we smooth -- with a boxcar function -- the AvPSD over 3 or 5 points depending on the evolutionary 
state of the star and the SNR of the spectra. From now on when we refer to the AvPSD we mean the 
smoothed AvPSD. An example of this spectrum is given in the top panel of Fig.~\ref{fig:AvPSD} for KIC~11244118. Because of the 
short length of the subseries, it is not necessary to interpolate the gaps in the data 
\citep[see for more details][]{2014A&A...568A..10G}, 
so we prefer the simplest possible analysis. 
However, we have verified that the results remain the same  when using series that were interpolated with the inpainting algorithm  \citep{2015A&A...574A..18P}.

To avoid subseries of low quality and to increase the SNR in the AvPSD we first remove all subseries with a duty cycle below 
$50\%$. Then, for each subseries of each star we compute the median of the flat noise at high 
frequencies, above 2 to 5 mHz depending on the frequency of maximum power of the star. From the statistical analysis of these medians, we reject those subseries in which the high-frequency noise is too high. We have verified that the selection of different high-frequency ranges does not affect the number of series retained. A detailed study of the rejected series show that there is an increase in the noise level when \emph{Kepler} has lost the fine pointing and the spacecraft is in `Coarse' pointing mode. This is particularly important in Q12 and Q16. A new revision of the KADACS software (Mathur, Bloemen, Garc\'\i a in preparation)  systematically removes those data points from the final corrected time series.
In Table~\ref{table:series} we summarize the number of 4-day subseries computed for each star,  the number of subseries finally retained for the calculation of the AvPSD, and the total number of short-cadence \emph{Kepler} quarters available. 

\begin{table}
\centering                         
\caption{KIC name of each star, number of initial 4-day subseries, number of finally averaged subseries after removing those 
with a low SNR,  and initial and final {\it Kepler} short-cadence quarters used in the analysis.}
\label{table:series}
\begin{tabular}{rccc} 
\hline\hline    
         KIC\;\;\;&$\#$Series& $ \#$Averaged  series &Qi - Qf\\
\hline         
3424541& $292$     &$ 270$&Q5-Q17\\
7799349& $292$     &$ 268$&Q5-Q17 \\
7940546& $ 244$     &$ 184$&Q7-Q17 \\
9812850& $292$     &$ 276$&Q5-Q17 \\
11244118& $292$     &$ 265$&Q5-Q17 \\
11717120& $292 $    &$ 276$&Q5-Q17 \\
\hline                       
\end{tabular}
\end{table}

\begin{figure}[!htb]
\begin{center}
\includegraphics[width=0.52\textwidth]{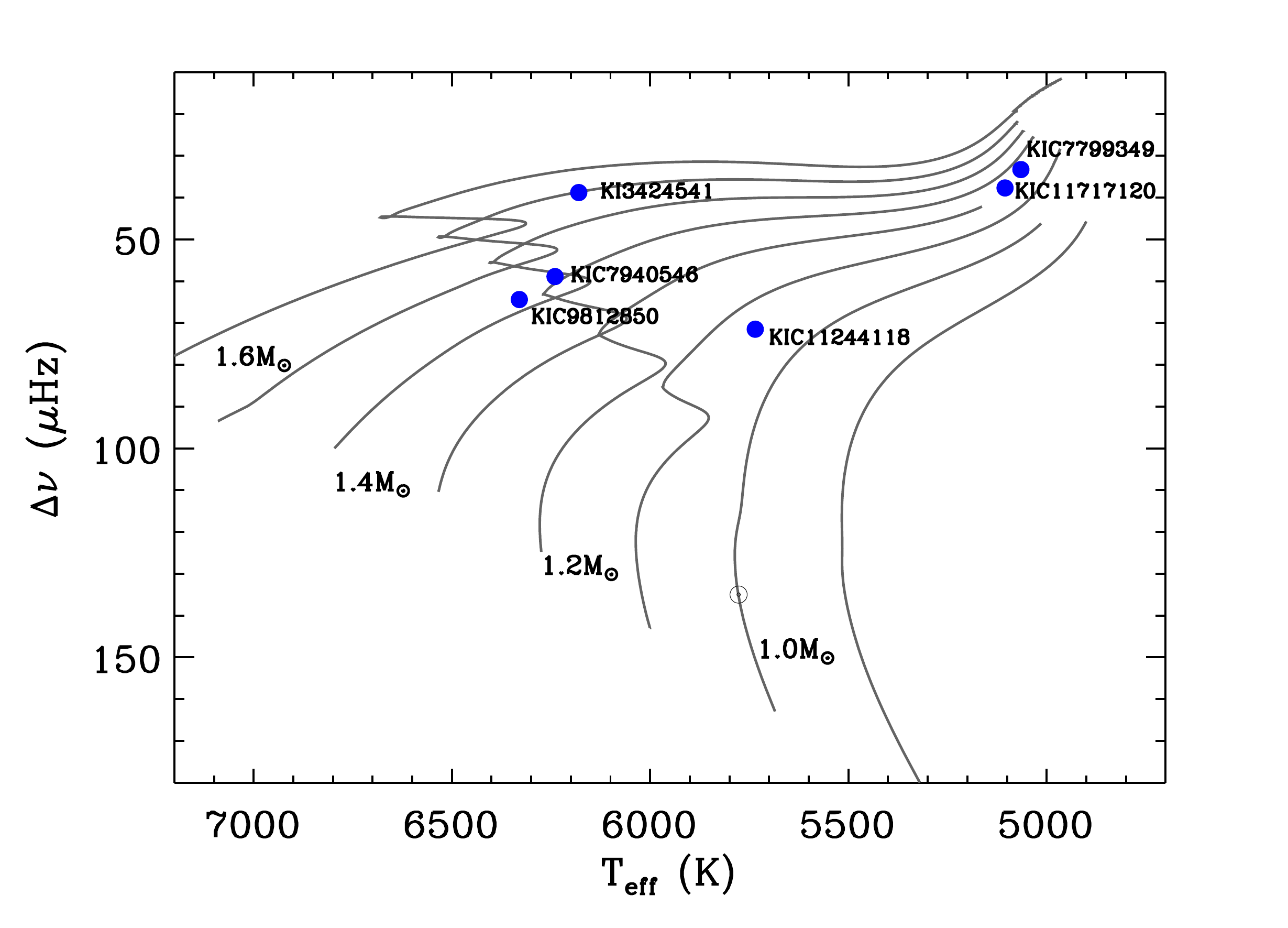}
\caption{Seismic HR diagram showing the position of the six stars analysed. The position of the Sun is indicated by the $\odot$ symbol. The evolutionary tracks are computed using ASTEC \citep {2008Ap&SS.316...13C}.}
\label{fig:hr}
\end{center}
\end{figure}

Fig.\ref{fig:hr} shows the place of the observed stars in a seismic HR diagram. The black lines are the evolution sequences computed using the Aarhus Stellar Evolution Code \citep[ASTEC][]{2008Ap&SS.316...13C} in a range of masses from
$M=0.9M_{\odot}$ to $M=1.6M_{\odot}$ in steps of $M=0.1M_{\odot}$ with a solar composition ($Z_{\odot} = 0.0246$). The target stars cover evolutionary stages from late main sequence stars to early red giants.

\section{The acoustic cut-off frequency}
\subsection{Observations}
The observed stellar power spectrum has different patterns in the eigenmode and pseudo-mode regions. In particular, the 
mean frequency separation, $\langle \Delta \nu \rangle$, between consecutive peaks is different below and above the cut-off frequency. In the first case, it 
corresponds to half the mean large frequency spacing, while above $\nucut$ it is the period of the interference 
pattern. In addition, a phase shift between both patterns appears in the transition region.
The frequency where the transition between the two regimes is observed corresponds to the cut-off frequency. Hence, we fit all the peaks in the spectrum (actually doublets of odd, $\ell=1$, 3, and even, $\ell=0$, 2, pairs of modes due to the 
small frequency resolution of $2.89\,\mu$Hz) from a few orders before $\nu_{\rm max}$ up to the highest visible peak in the 
spectrum. To take proper account of the underlying background contribution (due to convective movements at different 
scales, faculae, and magnetic/rotation signal), we  prefer to divide the AvPSD by the same spectrum heavily smoothed (see the 
red line in the top panel of Fig.~\ref{fig:AvPSD}) instead of using a theoretical model \citep[e.g.][]{2011ApJ...741..119M}. 
Although it has been demonstrated that a two-component model usually fits properly the background of stars \citep[see for more 
details][]{2014A&A...570A..41K}, the accuracy in the transition region between the eigenmode bump and the high-frequency 
region dominated by the HIPs is not properly described, which is the region we are interested in. Therefore, we use a simpler description of the background based on the observed spectrum itself. The length of this smoothing varies according to the evolutionary stage of the star. An example of the normalized resultant AvPSD (NAvPSD) is given in the bottom panel of Fig.~\ref{fig:AvPSD}.

A global fit to all the visible peaks in the spectrum is not possible for the following reasons: 1) it requires a huge amount of computation time owing to the high number of free parameters in the model, 2) the background is not completely flat around $\nu_{\rm max}$, and 3) the large amplitude dispersion of the peaks between $\nu_{\rm max}$ and the HIP pattern biases the results of the smallest peaks by overestimating their amplitudes. 
We therefore divide the spectrum into a few orders before $\nu_{\rm max}$ and the photon-noise dominated region into two regions. 
We arbitrarily choose a frequency where the amplitudes start to be very small and treat separately the power spectrum before and after this frequency. We denote by {\it p-mode region}  the low-frequency zone and {\it pseudo-modes region}  the other one. Several tests have been done in which the frequency separating these 
two regions were varied and in all cases the results remain the same.

\begin{figure}[!htb]
\begin{center}
\includegraphics[scale=0.4]{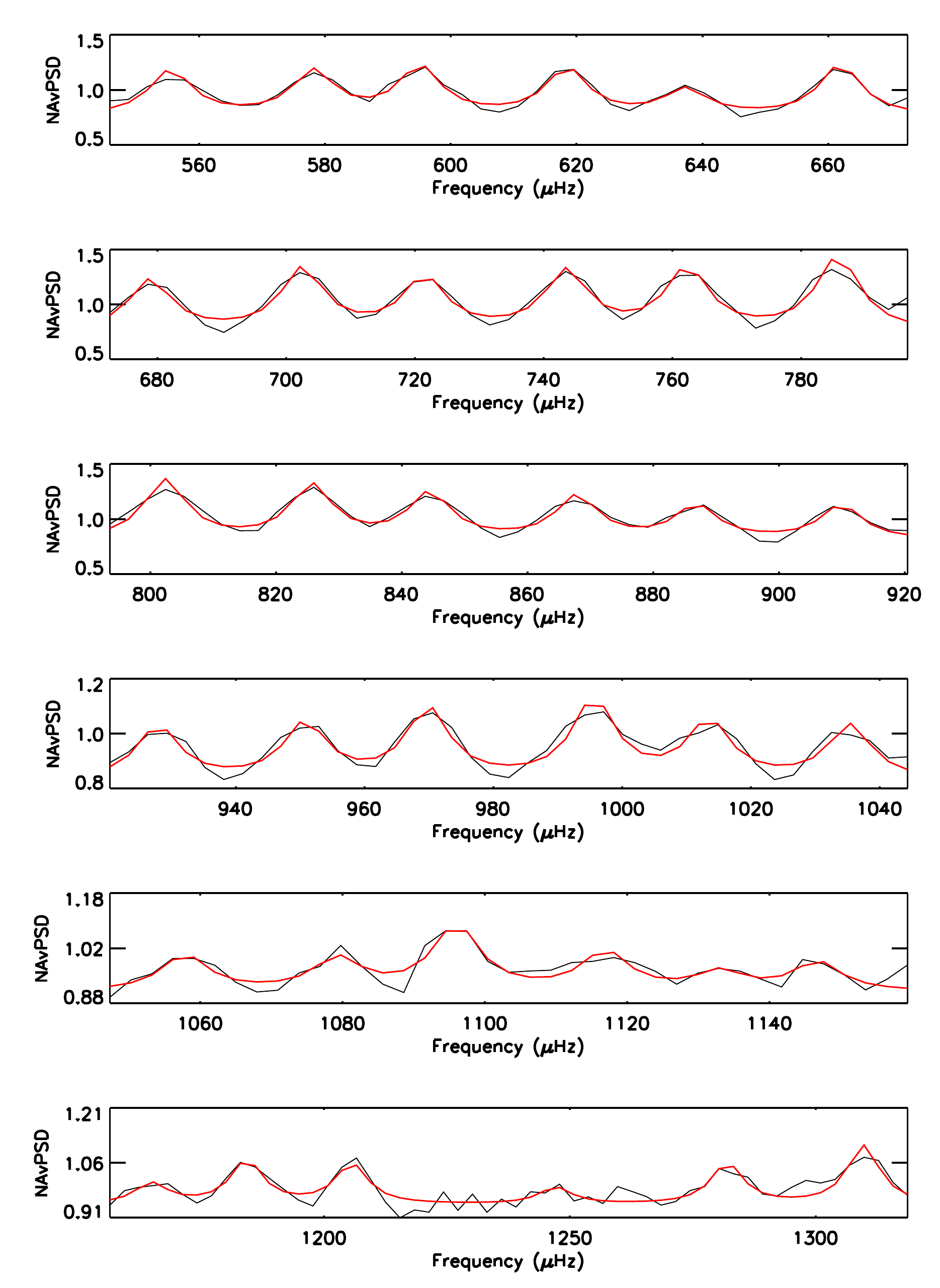}
\caption{Sections of the NAvPSD used to fit the eigenmode region by groups of six modes in KIC 3424541. The red line represents the final fit.}
\label{fig:pmodes}
\end{center}
\end{figure}

\begin{figure}[!htb]
\begin{center}
\includegraphics[scale=0.3,angle=90]{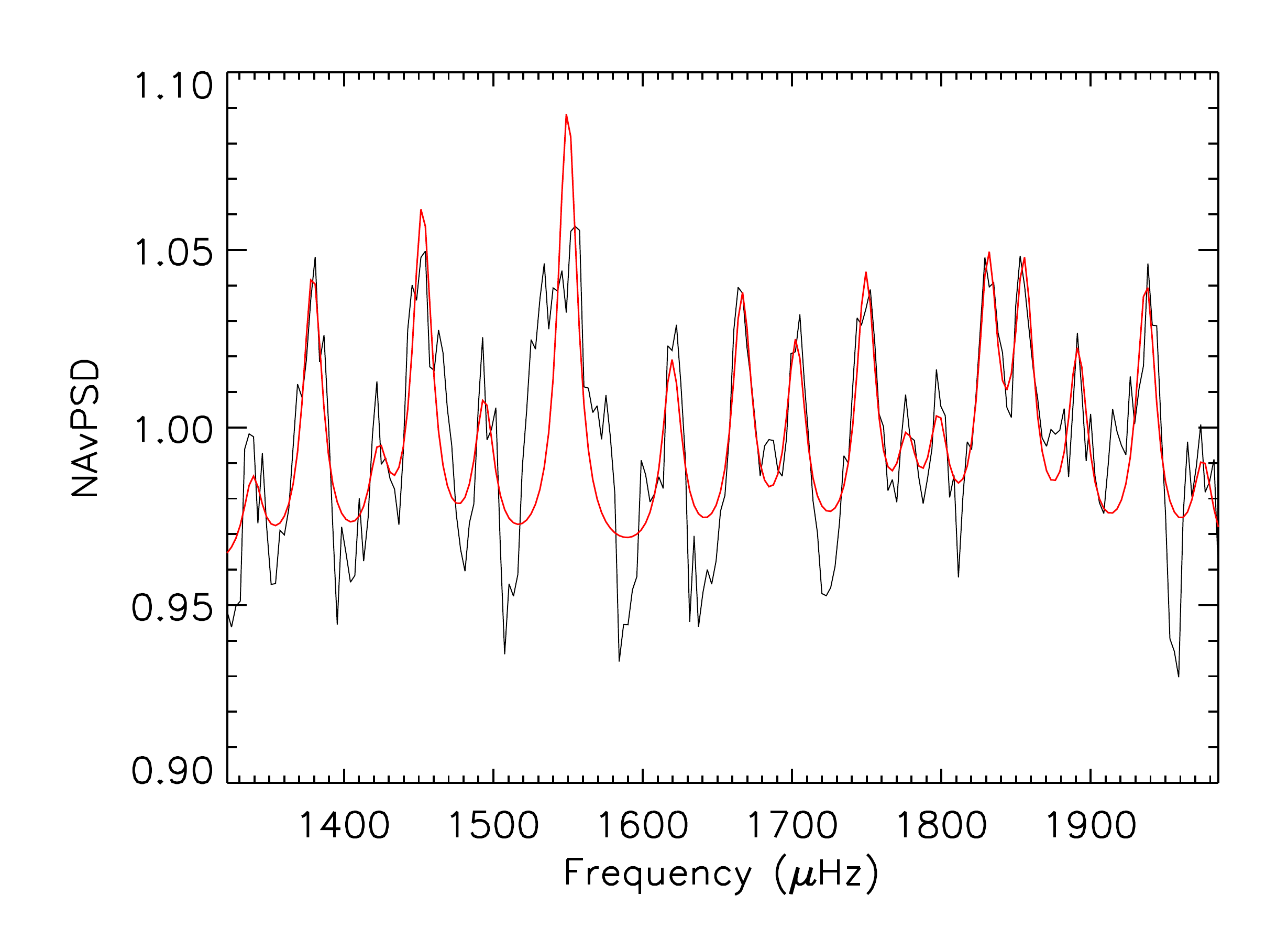}
\caption{Pseudo-mode region of the NAvPSD in KIC~3424541. The red line is the result of the fit. }
\label{fig:pseudos}
\end{center}
\end{figure}

In the p-mode region we fit groups of several peaks at a time for which the underlying background can be considered flat. This fit depends on the evolutionary stage of the star and the SNR. We have checked that fitting a different number of peaks in each group does not change the final result. Appendix A shows the analysis of all the stars, and in the caption of each figure we explicitly mention the number of peaks fitted together.
The pseudo-modes region can be fitted at once because 1) the background is flat (see the bottom panel of Fig.~\ref{fig:AvPSD}), 2) the range of the peak amplitudes is reduced, and 3) the number of free parameters is small. For both zones, p-mode and pseudo mode, a Lorentzian profile is used to model the peaks using a maximum likelihood estimator. 
It is important to note that we are interested only in the frequency of the centroids of the peaks, and for that a Lorentzian profile is a good approximation. In Figs.~\ref{fig:pmodes} and \ref{fig:pseudos} we show the results of the fits for KIC~3424541 in the p-mode and pseudo mode regions respectively. The figures for the other stars can be found in Appendix A.

\begin{figure}[!htb]
\begin{center}
\includegraphics[scale=0.3,angle=90]{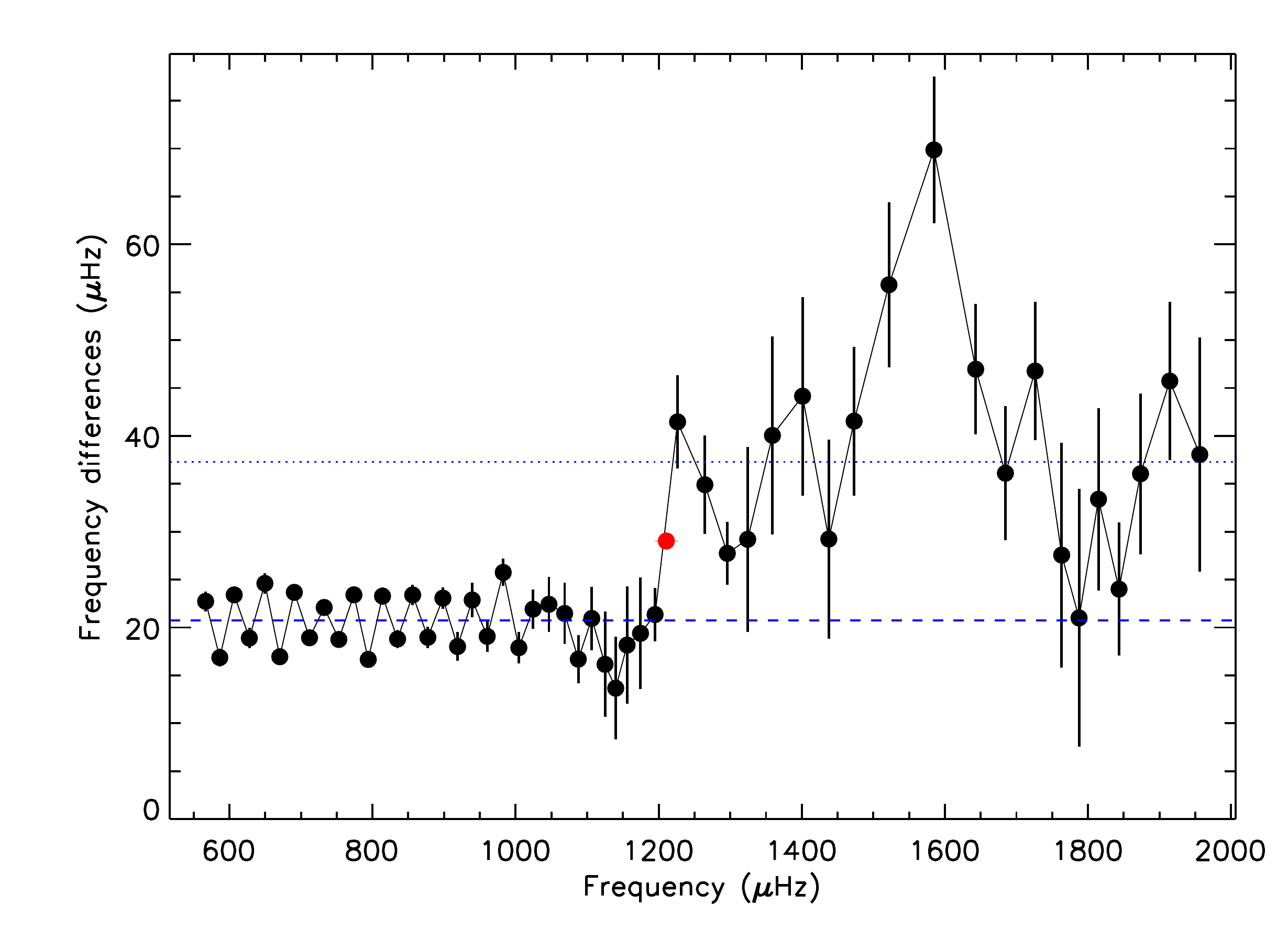}
\caption{Consecutive frequency differences of KIC~3424541. The blue dashed and dotted lines are respectively the weighted mean 
value of the frequency differences in the p-mode ($20.75\pm 0.20\,\mu$Hz) and pseudo-mode regions ($37.31\pm 1.58\,\mu$Hz) . The red symbol represents the estimated cut-off frequency, 
$\nucut= 1210.67\pm 15.70 \,\mu$Hz.}
\label{fig:freqdif}
\end{center}
\end{figure}

After fitting the two parts of the spectra, we compute the frequency differences of consecutive peaks:  
$\nu_{n}-\nu_{n-1}$. As an example, the resultant frequency differences for KIC~3424541 are plotted in Fig~\ref{fig:freqdif}. 

Two different regions are clearly visible in the frequency differences of KIC~3424541. Part of these differences -- those at 
lower frequencies -- are centred around $20.75\,\mu$Hz 
corresponding approximately
to half of the large spacing, $\dnumodes/2$ (see Table~\ref{tab:dnu}), owing to the alternation between odd and even modes. At a certain frequency, between 1200 and 1230 
$\mu$Hz, the frequency separations increase and remain roughly constant around $37.31\,\mu$Hz, although with a higher dispersion (see next section for further details). 
The observed acoustic cut-off frequency, $\nucut$, lies in the transition region where the differences jump from the p-mode 
region to the pseudo-mode zone. We define this position as the mean frequency between the two closest points to this transition, represented by a red symbol in Fig.~\ref{fig:freqdif} for KIC~3424541. The results of this analysis for the six stars studied here are given in 
Table~\ref{tab:nucut_numax}.

\begin{table*}[!htb]
\centering                 
\caption{\label{tab:dnu} KIC number of each star, effective temperature from \citet{2012MNRAS.423..122B} with an error bar of 84\,K, $\numax$ computed using the A2Z pipeline \citep{2010A&A...511A..46M}, $\nucut$, $\dnumodes$, $\Delta\nu_1$, $\Delta\nu_2$, $A_1$, and $A_2$ computed as explained in the text.     }  
\begin{tabular}{ccccccccc} 
\hline\hline    
     KIC\;\;\; & $\teff$   &$\numax$          &$\nucut$ & $\dnumodes$   & $\Delta\nu_1$  & $\Delta\nu_2$  & $A_1$ & $A_2$ \\
\hline     
Sun (GOLF)             & 5777 & 3097 $\pm$ 116 & $5106.41 \pm 61.53$ & $134.9\pm 0.8$ & $ 70.29 \pm  0.65$     & $143.77\pm 6.14$ & $ 0.018\pm 0.020$  & $ 0.008\pm 0.017$ \\
Sun (VIRGO)           & 5777 & 3097 $\pm$ 116 & $5106.41 \pm 61.53$ & $134.9\pm 0.8$ & $ 71.70\pm  0.56$     & $ 151.16\pm 14.33 $ & $ 0.056\pm0.038 $  & $ 0.010\pm0.024 $ \\
3424541& $	6180 $&$761.97\pm 39.10$ &$1210.67\pm 15.71$ & $38.8\pm  1.2$& $ 20.3\pm  2.1$     & $ 39.8\pm  1.5$  & $ 0.001\pm  0.018$ & $ 0.011\pm  0.016$ \\
7799349& $	5065 $&$568.14\pm 9.87$ &$ 895.47\pm 15.38$  & $33.3\pm  0.2$& $ 19.3\pm  2.5$     & $ 33.7\pm  0.9$ & $ 0.002\pm  0.017$ & $ 0.013\pm  0.016$\\
7940546& $	6240 $&$1099.57\pm 28.02$ &$1940.59\pm 20.71$ & $58.9\pm  0.4$& $ 30.3\pm  0.4$     & $ 48.9\pm  1.0$ & $ 0.021\pm  0.023$ & $ 0.020\pm  0.021$\\
9812850& $	6330 $&$1224.38\pm 52.21$&$1898.08\pm 24.29$ & $64.4\pm  0.5$& $ 31.8\pm  1.6$     & $ 54.0\pm  2.6$  & $ 0.009\pm  0.024$ & $ 0.018\pm  0.020$ \\
11244118& $	5735 $&$1384.07\pm 50.26$&$1968.69\pm 21.48$ & $ 71.5\pm  0.5$& $ 37.1\pm  2.0$     & $ 54.8\pm  3.4$  & $ 0.014\pm  0.029$ & $ 0.018\pm  0.025$ \\
11717120& $	5105 $&$582.96\pm 5.86$ &$ 958.38\pm 11.56 $ & $ 37.7\pm  0.3$& $ 16.6\pm  1.3$     & $ 29.9\pm  0.8$ & $ -0.003\pm  0.030$ & $ 0.030\pm  0.026$\\
\hline                       
\label{tab:nucut_numax}
\end{tabular}
\end{table*}

\subsection{Comparison between the observed and theoretical cut-off frequency\label{sec:models}}

\begin{figure}[!htb]
\begin{center}
\includegraphics[width=0.5\textwidth]{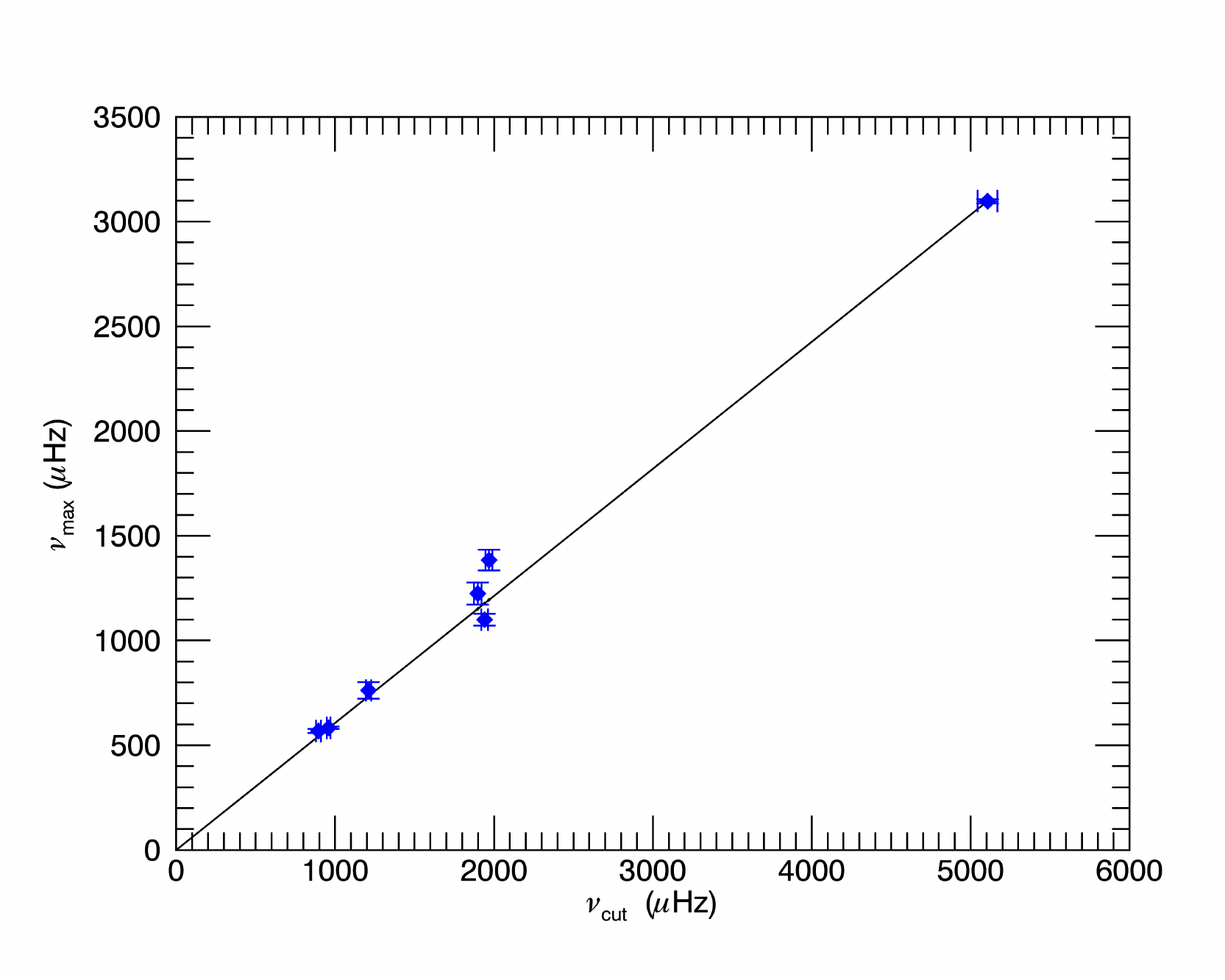}
\caption{Observed cut-off frequencies $\nucut$ versus maximum mode amplitude frequencies $\numax$. 
Blue points with errors are values for the six stars considered in this paper and the Sun (the rightmost point). 
A straight line through the Sun has been drawn for guidance.}
\label{fig:Th_numax_nucut}
\end{center}
\end{figure}

As noted in Sect.~\ref{sect:intro}, the cut-off frequency at the stellar surface 
scales approximately as $\nucut \propto  g \sqrt{\mu/\teff}$, suggesting that we may compare the observational cut-off frequency with that 
computed from the spectroscopic parameters. However, for our set of stars, the spectroscopic $\log g$ has large uncertainties
\citep[see][]{2012MNRAS.423..122B} and a direct test is not possible. Alternatively, it has been demonstrated empirically that 
the frequency of maximum mode amplitude $\numax$ follows a similar scale relation \citep[see][]{2003PASA...20..203B}. 
As  can be seen in Fig. \ref{fig:Th_numax_nucut}, the observational data show such a linear relation 
between $\numax$ and $\nucut$, although some stars deviate 2$\sigma$ from it. 
In fact, from a theoretical point of view, one does not expect this linear relation to be accurate enough at the 
level of the observational errors. As shown below, values of $\nucut$ 
can deviate by as much as $10\%$ from the scale relation $\nucut \propto  g \sqrt{\mu/\teff}$
for representative models of our stars. On the other hand, 
following \cite{2008A&A...485..813C},
where theoretical values of $\numax$ based on stochastic mode excitation were computed, one can find
departures from the scale relation 
as large as $25\%$ for a model of a star with $M=1.3M_{\odot}$ and $\numax\sim 1500\,\mu$Hz.

We can gain a deeper insight by comparing the observational and theoretical results in the $\nucut$--$\Delta\nu$ plane. 
For the theoretical calculations we use a set of models computed with the CESAM code \citep{2008Ap&SS.316...61M}. 
These correspond to evolutionary tracks from the zero-age main sequence up to the red-giant phase, with masses between 
$M=0.9M_{\odot}$ and $M=2.0M_{\odot}$, and 
helium abundances between $Y=0.25$ and $Y=0.28$. Models with and without overshooting were considered.
Other parameters were fixed to standard values, such as the metallicity to $Z=0.02$. 
Finally, the CEFF equation of state \citep{1992A&ARv...4..267C} and a $T(\tau)$ relation derived from a solar atmosphere model were used. 

In theoretical computations, a standard upper boundary condition for eigenmodes is to impose in the uppermost layer the simple 
adiabatic evanescent solution for an isothermal atmosphere. In this case eigenmodes turn to have frequencies below a cut-off 
frequency given by $\omegacut=2\pi \nucut=c/2H$, where $H$ is the density scale height
(this is strictly speaking the cut-off frequency for radial 
oscillations, which is accurate for low-degree modes). To better fulfil the quasi-isothermal requirement, the boundary condition should be placed near the minimum temperature. For the Sun this corresponds to an optical depth of about $\log\tau=-4$.  Another advantage of using such a low optical depth is that at this position the oscillations are not too far from being adiabatic, at least compared to the photosphere. In our computations we have used this value for all the models as the uppermost point, 
but for some red-giant stars we found that the maximum value of $\omegacut(r)$ in the atmosphere can be located at higher 
optical depths; hence, we have taken that maximum value as representative of the observed $\omegacut$. 
From an interpolation to the solar mass and radius (with $Y=0.25$ fixed) we obtain
from our set of models a value of $\nucut$ = 5125 $\mu$Hz for the Sun, which is in agreement with the observed one: 
$\nucut =5106.41\pm 61.53\mu\,$Hz \citep{2006ApJ...646.1398J}. 
If model S \citep{1996Sci...272.1286C} is considered, the differences are a little larger, about $4\%$. 
It is important to remember that the solar $\nucut$ changes with the magnetic activity cycle \citep{2011ApJ...743...99J}; thus, 
some additional dispersion in the observed stellar values could be due to that effect.

For the large separation, the theoretical and observed values can be computed in the same way. Specifically, we have computed 
the large separation $\dnumodes$ 
from a polynomial fit of the form $(\nu - \nu_0)/(n -n_0) = \sum_i a_i P_i (x)$, where $n$ is the radial order and 
$\nu_0$ is the radial frequency closest to $\numax$ with radial order $n_0$.
For the models, $\nu_0$ is estimated by assuming a linear relation to the cut-off frequency. 
The frequencies considered are those of the $\ell=0$ modes with radial orders $n=n_0\pm 4$. $P_i(x)$ is the Legendre 
polynomial of degree $i$, and $x$ is the frequency normalized to the interval $[-1,1]$. A third order polynomial has been used. 
From the Tasoul equation we expect that $a_0\equiv \dnumodes\simeq \Delta \nu$, whereas other terms will mostly contain upper-layer information. We use only radial oscillations because, for evolved stars, the mixed character of some modes can introduce 
complications. 
For the Sun we obtain $\dnumodes=134.9\,\mu$Hz from the observations and $\Delta\nu_{\rm modes}=136.0\,\mu$Hz from model~S.

\begin{figure}
\begin{center}
\includegraphics[width=0.45\textwidth]{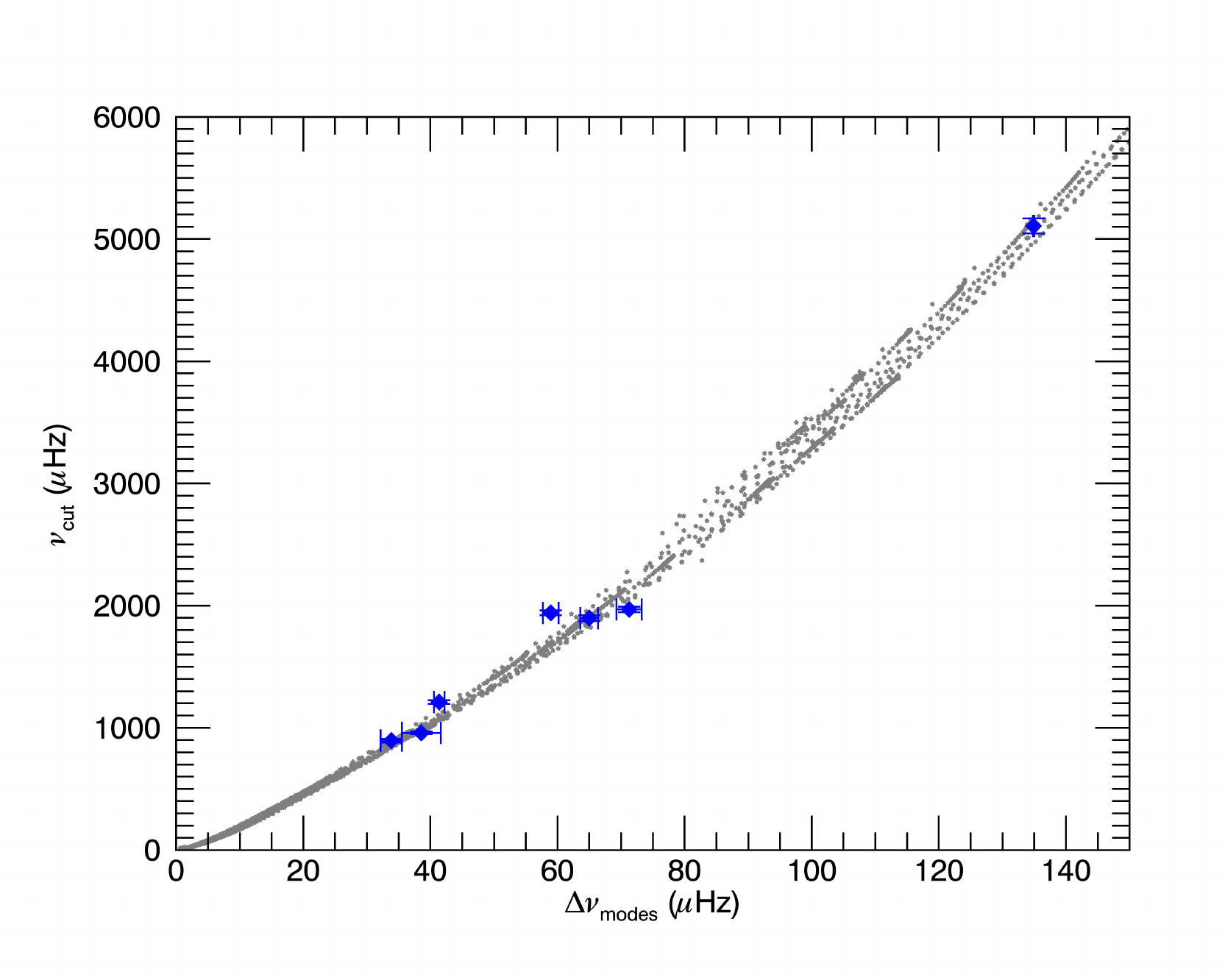}
\caption{$\nucut$ versus $\dnumodes$ for our set of theoretical models (grey dots). Blue points correspond to the observed 
stars, including the Sun (upper-right point).}
\label{fig:deltanu_nucut}
\end{center}
\end{figure}

Figure~\ref{fig:deltanu_nucut} shows $\nucut$ against the large separation $\dnumodes$ for our set of models and the 
observed stars. At first glance, there is rough agreement between the observed and theoretical calculations. but for the group of stars with $\nucut \sim$2000 $\mu$Hz 
it seems hard to explain the dispersion in their $\dnumodes$ values.

\begin{figure}
\begin{center}
\includegraphics[scale=0.4]{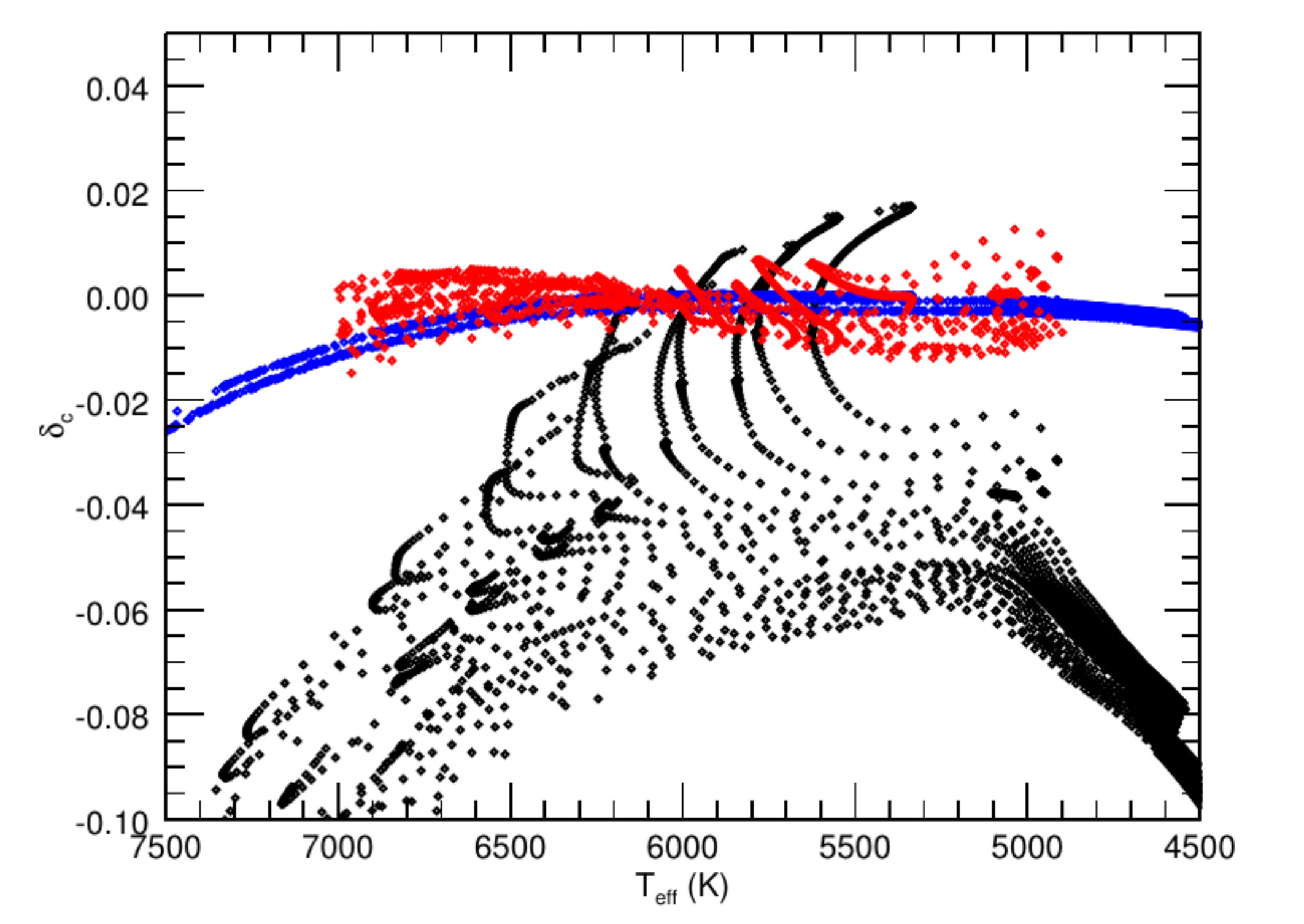}
\caption{Relative differences $\delta_{\rm c}$ as a function of $\teff$ (black dots). Blue points are the same relative 
differences but considering the isothermal cut-off frequency $\omega_{\rm a}$ instead of $\omega_{\rm c}$. Red points are the residuals after correcting by the polynomial fit indicated
in the text. The fit is limited to models with $\teff<7000$ K and $\nucut>500$ $\mu$Hz.}
\label{fig:nucut}
\end{center}
\end{figure}

We can proceed forward by taking into account the values of $\teff$. First, given the approximated scale relation for $\nucut$,
we compute the residuals:
\begin{equation}
\delta_{\rm c} =1 - \frac{\nucut^0}{\nucut} \, \frac{ g\sqrt{\mu/\teff} } {\left(g\sqrt{\mu/\teff}\right)_0 }
\; ,
\label{eq:deltac}
\end{equation}
where $\nucut^0$ = 5079.5 $\mu$Hz is a reference cut-off frequency, corresponding to a model in our dataset close to the 
Sun ($M=1M_{\odot}$, $R=1.004R_{\odot}$, $Y=0.25$, $\teff^0$ = 5799.2 K).
The  $\teff$ values are shown in Fig. \ref{fig:nucut}.
As seen in the figure, the differences can be as large as $|\delta_{\rm c}| \approx 10\%$. 
In fact, the scale relation $\nucut \propto  g \sqrt{\mu/\teff}$ is derived theoretically by
approximating the density scale height, $H$, by the pressure scale height, $H_p$, which is strictly valid in an
isothermal atmosphere, and further by taking the equation of state of a mono-atomic ideal gas. 
In Fig. \ref{fig:nucut}, the blue points correspond to the relative differences 
$\delta_{\rm a}$ obtained by replacing in Eq. \ref{eq:deltac} $\omegacut=c/2H$ by $\omega_{\rm a}=c/2H_p$. 
Thus, the first condition introduces the highest departures from the scale relation, which was expected because in the  
range of $\teff$ considered, the gas is mainly mono-atomic at the surface.

Nevertheless, as seen in Fig. \ref{fig:nucut}, $\delta_{\rm c}$ is mainly a function of $\teff$. Hence, the scale relation can be improved if we subtract a polynomial fit from $\delta_{\rm c}$. Only models with $\teff<7000\,$K and $\nucut>500$ $\mu$Hz have been considered in that fit because this range includes all the stars used in the present study. For a better fit we also consider the dependence of $\delta_{\rm c}$ on $\nucut$. In particular, the red points in Fig.~\ref{fig:nucut} are obtained by replacing $\nucut$ in Eq.~\ref{eq:deltac}  by:
\begin{eqnarray}
\nucut^* = \nucut \left[ 1 - (0.0004 -0.0057 x - 1.0684x^2 +0.0543y +\right. \nonumber \\
\left. + 0.1265 x y -0.7134 x^2 y) \right] \frac{\nucut (\odot)} {\nucut^* (\odot)},
\label{eq:nucef}
\end{eqnarray}
where $x=\teff/\teff^0-1$ and $y=\nucut/\nucut^0-1$. 
The standard deviation for these residuals is $0.3\%$. Hence, the modified cut-off frequencies, that we calibrated
on the Sun deviate by $1\sigma=0.3\%$ from the scaling relation, at least for our set of models. 

\begin{figure}
\begin{center}
\includegraphics[scale=0.4]{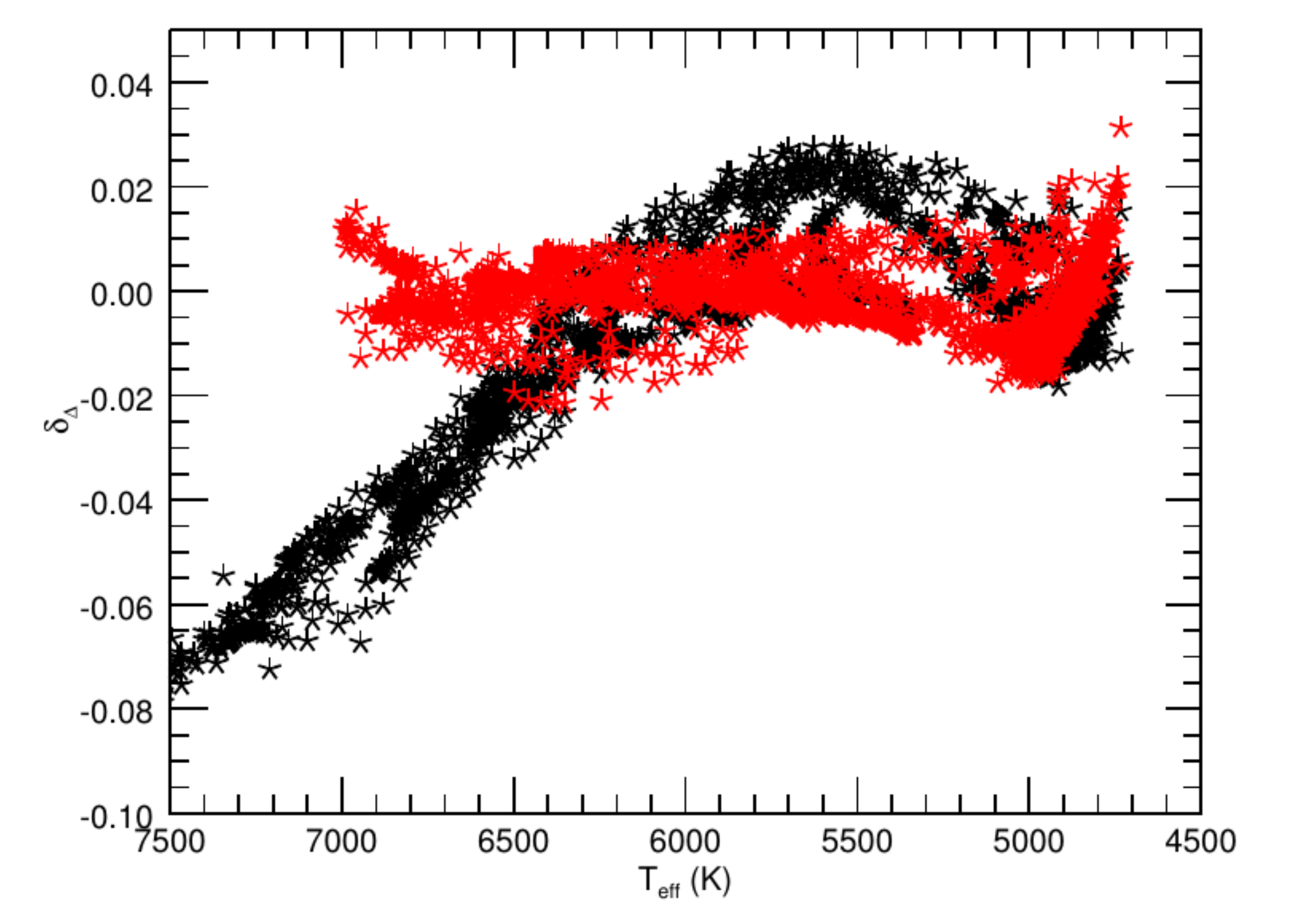}
\caption{Relative differences $\delta_{\Delta}$ versus $\teff$ for the set of models indicated
in the text (black dots). The red points are the residuals after the polynomial fit. This fit is limited to models with $\teff<7000\,$K and $\Delta\nu>10\mu$Hz.}
\label{fig:dnu_scale1}
\end{center}
\end{figure}

\begin{table*}[!ht]
\centering                         
\caption{Values of $\log g$ derived from spectroscopy, $\numax$ and $\nucut$ for each star in our sample.
The mean density $\bar\rho$ is in g/cm$^3$ and the large separation $\dnumodes$ in $\mu$Hz. See text for details.}
\label{table:parameters}
\begin{tabular}{r c c c c c } 
\hline\hline    
  KIC\;\;\;&$\log g_{\rm{sp}}$&$\log g_{\numax}$ & $\log g_{\nucut}$  &$\dnumodes$  & $\bar\rho$  \\
\hline         
3424541&$3.50\pm 0.08 $ &$3.84\pm 0.03$& $3.86\pm 0.02$&$ 38.8\pm  1.2$& $0.116\pm 0.007$ \\
7799349&$3.71\pm 0.08 $ &$3.67\pm 0.02$& $3.68\pm 0.02$&$ 33.3\pm  0.2$& $0.085\pm 0.001$ \\
7940546&$4.11\pm 0.08 $ &$4.01\pm 0.02$& $4.06\pm 0.02$&$ 58.9\pm  0.4$& $0.270\pm 0.004$ \\
9812850&$4.16\pm 0.08 $ &$4.06\pm 0.03$& $4.06\pm 0.02$&$ 64.4\pm  0.5$& $0.327\pm 0.005$ \\
11244118&$4.23\pm 0.08 $ &$4.09\pm 0.03$& $4.04\pm 0.02$&$ 71.5\pm  0.5$& $0.388\pm 0.005$ \\
11717120&$3.80\pm 0.08 $ &$3.69\pm 0.02$& $3.71\pm 0.02$&$ 37.7\pm  0.3$& $0.109\pm 0.002$ \\
\hline                       
\end{tabular}
\end{table*}
In a similar way, we compute the deviation of $\dnumodes$ from its expected scale relation, namely: 
\begin{equation}
\delta_{\Delta} =1 - \frac{(\dnumodes)^0}{\dnumodes}  \sqrt{ \frac{M/M_0}{(R/R_0)^3}}
\label{eq:delta_delta}
\end{equation}
where $(\dnumodes)^0$ is the value for our reference model, the same as the one used in $\nucut$.
This quantity is shown in Fig.~\ref{fig:dnu_scale1} for our set of models.

From Fig.~\ref{fig:dnu_scale1} we can determine that the differences between the large separation computed from p-modes and the scaling relation can be as large as $5\%$ for
solar-like pulsators with $\teff < 7000\,$K. 
However, as shown in this figure, the differences $\delta_{\Delta}$ depend mainly on $\teff$. This was previously noted by \cite{2011ApJ...743..161W} and allows for a correction to $\dnumodes$ in order to 
get better a estimate of the mean density. Here we will use a fit similar to that considered for $\nucut$. As in the previous case, we have considered only those models with $\teff<7000$ K and $\dnumodes>10$ $\mu$Hz, which include all the stars used in the present work. The red points in Fig.~\ref{fig:dnu_scale1} correspond to the residuals
obtained after replacing $\dnumodes$ in Eq.~\ref{eq:delta_delta}  by:
\begin{eqnarray}
\dnumodes^* = \dnumodes \left[ 1 -(0.0023 -0.1123 x - 0.7309x^2 - \right. \nonumber \\
{\left.  -0.0299 z -0.0067 x z + 1.2461 x^2 z ) \right.] } \frac{\dnumodes (\odot)} {(\dnumodes)^* (\odot)}
\label{eq:dnuef}
\end{eqnarray}
where $x=\teff/\teff^0-1$ and $z=\dnumodes/(\dnumodes)^0-1$ with $(\dnumodes)^0=135.4$ $\mu$Hz. 
The standard deviation for these residuals is $0.6\%$.
Given the differences between the 
observed and theoretical values for the Sun, we have calibrated Eq.~\ref{eq:dnuef} to the observed solar value.

\begin{figure}
\begin{center}
\includegraphics[width=0.45\textwidth]{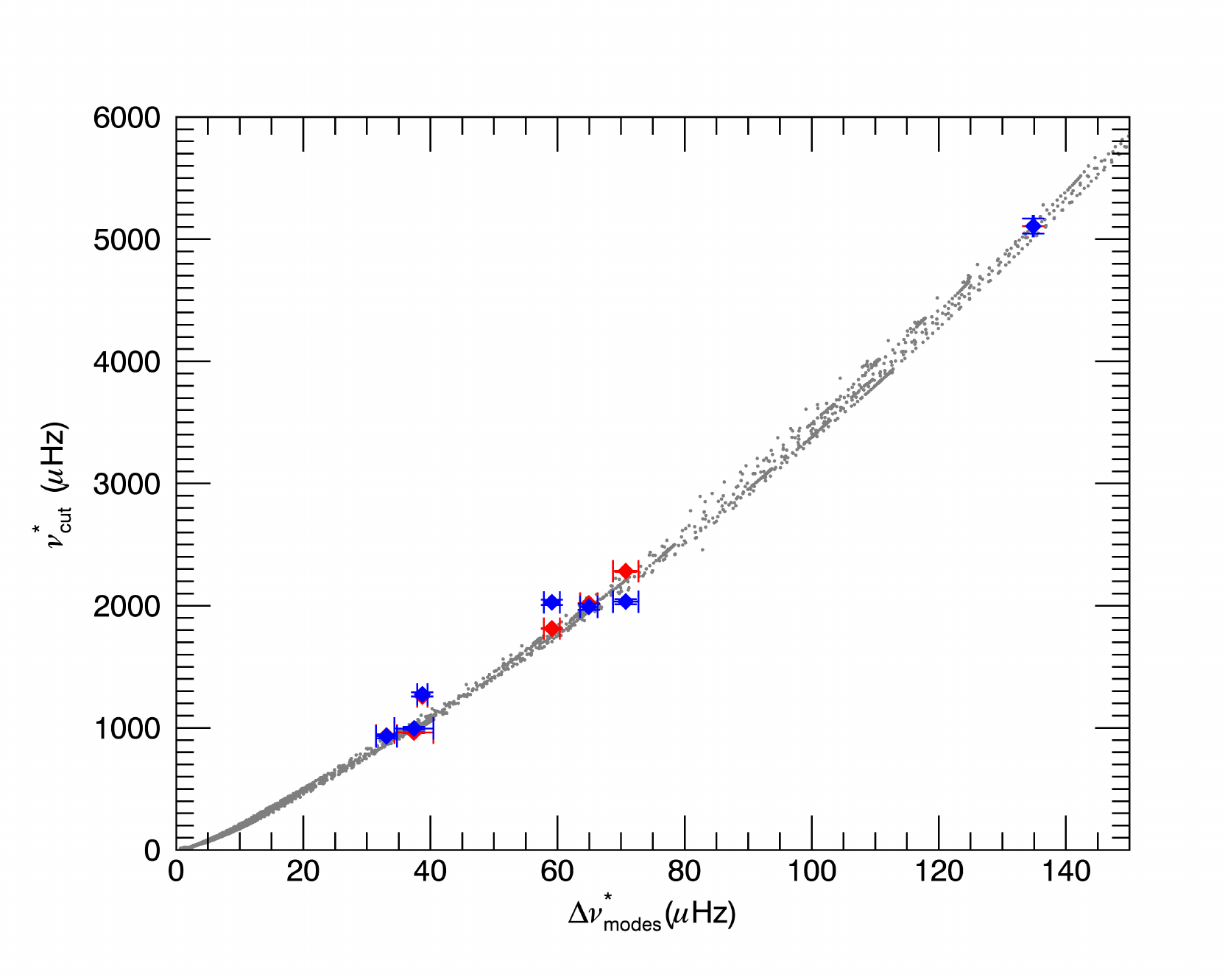}
\caption{Effective $\nucut^*$ versus the effective $\Delta \nu_{\mathrm{modes}}^*$ for our set of models (grey dots) and for 
the observed stars (blue dots). Red points are observed values of the frequencies of maximum amplitude properly scaled,
$(\nucut/\numax)_{\odot}\numax$.}
\label{fig:dnu_nucut}
\end{center}
\end{figure}

Figure~\ref{fig:dnu_nucut} shows $\nucut^*$ against $\Delta \nu_{\mathrm{modes}}^*$ for our set of models (black points) and the observed stars including the Sun (blue points). Whereas the scattering in the theoretical values are substantially reduced, the observational ones are not. Hence, if the error estimates are correct, we must conclude that the observational cut-off frequencies do not completely agree with the theoretical estimates. 
Moreover, the red points in Fig.~\ref{fig:dnu_nucut} are proportional to the observed values of the frequency of maximum 
amplitude scaled to the Sun, $(\nucut /\numax)_{\odot}\numax$. 
For our limited set of stars, $\numax$ follows the scaling relation better than $\nucut^*$. 
This is a little surprising because, as noted before, theoretically, we would expect the opposite, even more so when 
the effective $\nucut^*$ is used. 

This result can be understood in terms of the inferred surface gravity since we can use either $\numax$ or $\nucut$ to estimate $\log g$. The results are summarized in Table~\ref{table:parameters} and compared to the spectroscopic $\log g_{\rm{sp}}$. To estimate the errors in $\log g_{\numax}$ we have considered only the observational errors 
in $\numax$ and $\teff$, while for $\log g_{\nucut}$ we have included the 1-$\sigma$ value of $0.3\%$ found in the scaling relation plus an error in $\mu_{\rm s}$ derived from assuming an unknown composition with standard stellar values. In particular, we have taken ranges of $Y=[0.24,0.28]$ and $Z=[0.01,0.03]$ for the helium abundance and the metallicity respectively and thus estimated the error in the mean molecular weight $\mu_{\rm s}$ by some $5\%$. As expected, the values obtained from $\log g_{\numax}$ are quite similar 
to those reported by \citet{2012MNRAS.423..122B}
because they are based on the same global parameters with very similar values.
The values derived from $\numax$ and $\nucut$ are also much closer to each other than those derived spectroscopically, which, as mentioned before, have large observational uncertainties.

\section{The HIP region}

\subsection{Observations}
Using Sun-as-a-star observations from the GOLF instrument on board SoHO, \citet{GarPal1998} uncovered the existence of a sinusoidal pattern of peaks above $\nucut$ 
as the result of the interference between two components of a travel wave generated on the front side of the Sun with a frequency $\nu \geq \nucut$,  where the inward component returns to the visible side after a partial reflection on the far side of the Sun \citep[see Fig.~3 in][]{GarPal1998}. Therefore, the frequency spacing of $\sim$70 $\mu$Hz found in the Sun corresponded to the time delay between the direct emitted wave and that coming from the back of the Sun 
(corresponding to waves behaving like low-degree modes), i.e.\ a delay of four times the acoustic radius of the 
Sun ($\sim$ 3600s). This value is roughly half of the large frequency spacing of the star and we will call it 
$\Delta\nu_1$. \citet{GarPal1998} also speculated that, above a given frequency, a second pattern should become visible with a 
double frequency spacing (close to the large frequency separation), $\Delta\nu_2 \sim$ 140 $\mu$Hz. This pattern \citep[see 
for a theoretical description][]{1993ASPC...42...15K} -- usually visible in imaged instruments \citep{1991ApJ...373..308D} -- 
corresponds to the interference between outward emitted waves and the inward components that arrive at the visible side of the 
Sun after the refraction at the inner turning point (non-radial waves). 

In order to get a global estimate of the frequency of the interference patterns in the HIP region of our sample of stars, two sine waves are fitted above the cut-off frequency. The amplitudes and the frequencies of both sine waves are summarized in Table~\ref{tab:dnu}.  We have also re-analysed the GOLF and VIRGO data following the same procedure but at a higher frequency range, between 7 and 8 mHz, compared to the original analyses performed by \citet{GarPal1998} and \citet{2006ApJ...646.1398J}. In this way, we have been able to obtain the second periodicity at $\sim$ 140 $\mu$Hz (see Table~\ref{tab:dnu}).

We tried fitting various functions and opted for the simplest one. Indeed, the amplitude of the interference patterns decreases with frequency in a way close to an exponential decrease. The additional parameters required give more unstable fits with a heavy dependence on the guess parameters.  Because we obtained the same qualitative results leading to the same classification of stars, we preferred this approach.
In future studies we will look for a better function for the fit -- i.e.\ one more suited to the observations -- in a larger set of stars. We are already working in this direction but this investigation is beyond of the scope of this paper.

\begin{figure}[!htb]
\begin{center}
\includegraphics[scale=0.3,angle=90]{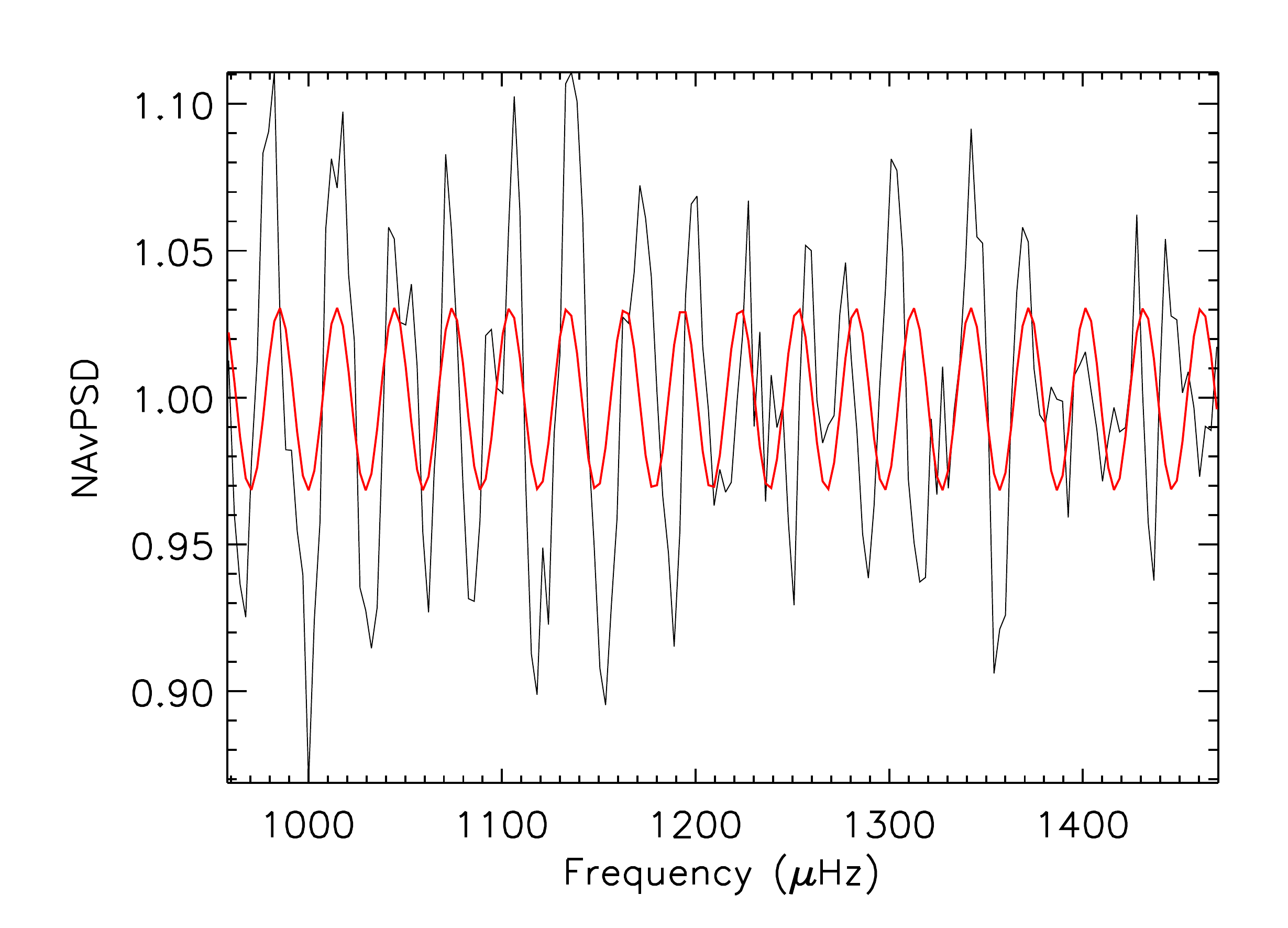}
\includegraphics[scale=0.3,angle=90]{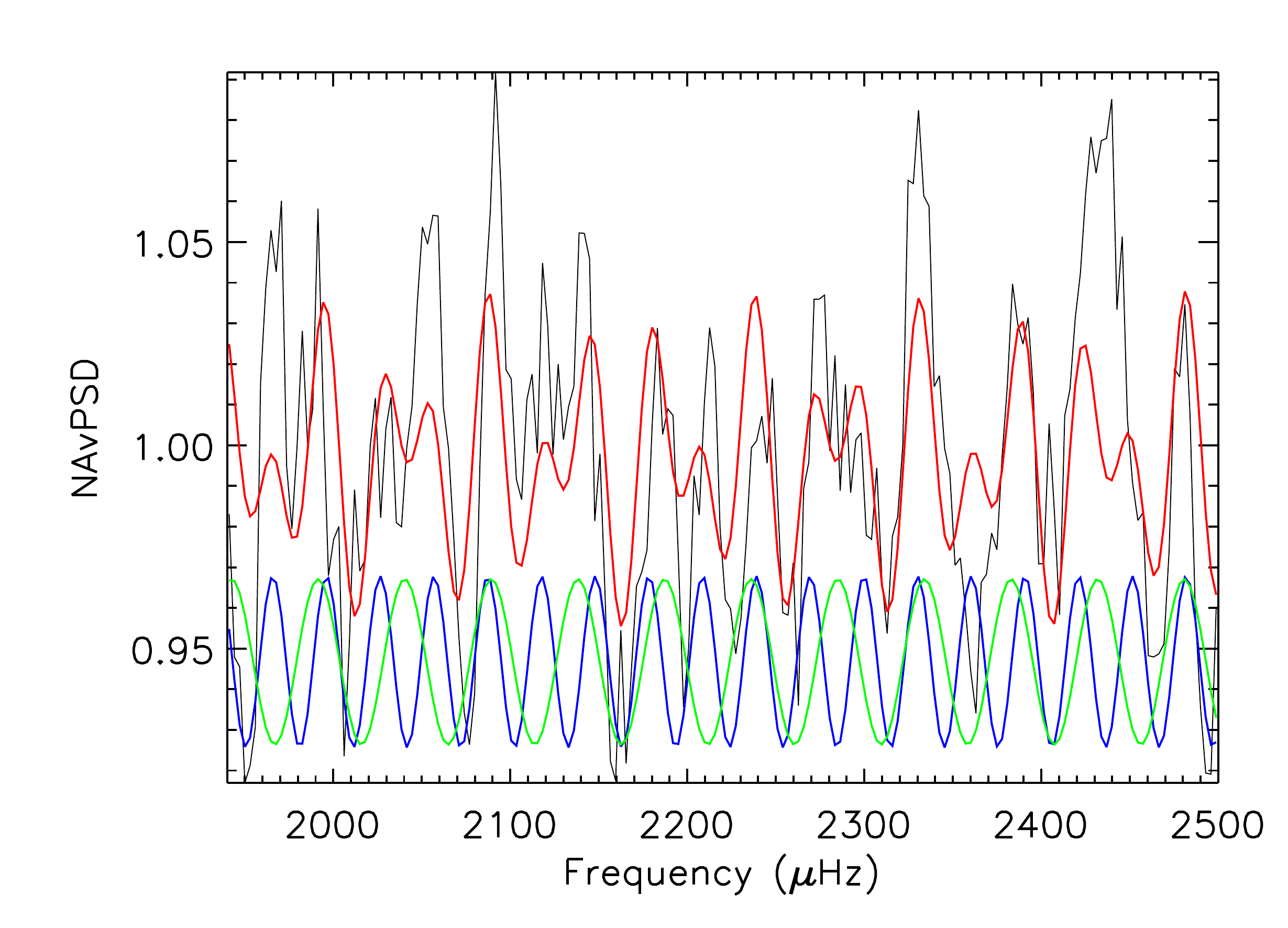}
\caption{HIPs region of KIC~11717120 (top), where the HIP pattern is dominated by the sine wave (red line) owing to the interference between direct and refracted waves, and of KIC~7940546 (bottom), where the two HIP patterns can be determined. The blue and green line in the bottom panel correspond to the two fitted sine waves with frequencies $\Delta\nu_1$ and $\Delta\nu_2$ respectively. 
The green and blue lines have been shifted by 0.05 for clarity. The red line is the actual fit.}
\label{fig:794}
\end{center}
\end{figure}

According to the fitted amplitudes, we can classify the stars into two groups: those where the two amplitudes of the sine waves, $A_1$ and $A_2$, are similar (KIC~7940546, KIC~9812850, KIC~11244118), and those where $A_2$ is much larger than $A_1$ (KIC~3424541, KIC~7799349, KIC~11717120). An example of each group of stars is given in Fig.~\ref{fig:794} for stars KIC~7940546 and KIC~11717120. It is important to notice that the error bars of the fitted amplitudes are large  because the actual amplitudes of the interference patterns decrease in a quasi-exponential way that has not been taken into account here. In this pioneering work, we have favoured the fits with constant amplitudes to avoid adding more unknown parameters to the fit.

The stars of the first group have larger $\dnumodes$ than the other three stars, which implies a correlation with the evolutionary state of the stars. Starting with the Sun on the main sequence, the HIP pattern is dominated by interference waves partially reflected at the back of the Sun (for the solar case $A_1$ is much bigger than $A_2$).
When the stars evolve, the HIP pattern due to the interference of direct with refracted waves in the stellar interior is more and more visible and becomes dominant for evolved RGB stars with $\dnumodes$ below $\sim$40 $\mu$Hz. 

\subsection{HIPs and stellar evolution}

In order to reproduce theoretically the periodic signals expected from the pseudo-mode spectrum we will follow the 
interpretation given by \cite{1993ASPC...42...15K} and \cite{GarPal1998} and assume that waves are excited isotropically at a 
point very close to the photosphere. The observed spectrum of the pseudo modes is then interpreted as
an interference pattern between outgoing and ingoing components, possibly including successive surface reflections.
In what follows we summarize the basic concepts and apply them to stars at different evolutionary stages.

Let us start by using a gravito--acoustic ray theory approximation with the 
following dispersion relation  \citep[see e.g.][]{1986hmps.conf..117G,1993afd..conf..399G}:
\begin{equation}
k^2 = k_{\rm r}^2 + k_{\rm h}^2 = \frac{1}{c^2} \left(\omega^2 -\omega_c^2 \right) 
+ \frac{N^2}{\omega^2} k_{\rm h}^2 
\, ,
\label{edisp}
\end{equation}
where $k_{\rm r}$ is the radial component of the wave number, $k_{\rm h}^2 = L^2/r^2$, $L=l(l+1)$ the horizontal component, 
$N$ is the buoyancy frequency, $\omega_c$ is a generalized cut-off frequency given by
\begin{equation}
\omega_c^2 = \frac{c^2}{4H^2} \left( 1 - 2 \frac{dH}{dr} \right)
\, ,
\end{equation}
and $H$ is the density scale height.
In the ray approximation the turning points are given by $k_{\rm r}=0$, whereas 
the ray path is determined by the group velocity. For the 
spherically symmetric case the rays are contained in a plane with a path given by $d \theta/d r = v_{\theta}/(r v_r)\,$. 
Here, $r$ and $\theta$ are the usual polar coordinates and the group velocity 
${\bf v}_{\rm g} = (v_r,v_{\theta})= ( \partial \omega/\partial k_r , \partial \omega/\partial k_{\rm h})$
has the following components:
\begin{equation}
{\bf v}_{\rm g} =
\left( \,\frac{k_r \omega^3 c^2} {\omega^4 - k_h^2 c^2 N^2} , k_{\rm h} \, \omega \, c^2 
\left[ \frac{\omega^2 - N^2}{\omega^4 -k_{\rm h}^2 c^2 N^2} \right] \,\right)
\, .
\end{equation}

According to the ray theory, the general solution of the wave equation can be expressed as a superposition of rays 
of the form $\psi_{\omega} (r,t) = A_{\omega}(r) \exp [i (\omega t \pm  \int_s k \, d s)]$,
where the two signs corresponds to outgoing and ingoing waves and
the integral is computed over the ray path $s$, including additional reflections where appropriate. 
For every reflection, a constant phase shift must be introduced.
In particular, for a ray travelling from an inner turning point $r_1$ to an outer turning point $r_2$, this integral can 
be expressed as 
\begin{equation}
\frac{\omega}{2\Delta\nu_{\omega,\ell}} \equiv 
\int_s k \, d s = \int_{r_1}^{r_2}  k \frac{v_{\rm g}}{v_r}\, dr - \pi/4
\; .
\end{equation}
For low-degree acoustic waves $\omega/(2\Delta\nu_{\omega,\ell}) 
\simeq \omega \int_0^R (1/c) dr -(1+\ell) \pi/2 \,$, that is,
$\Delta\nu_{\omega,\ell} \approx \Delta\nu\,$.

Considering a wave excited very close to the surface with an outgoing component 
of amplitude $A_{\omega\ell}$ and an ingoing component that emerges at the surface with an amplitude 
$A'_{\omega\ell}$ after its first internal traversal, and with an amplitude of $A'_{\omega,\ell}R(\omega)$ after
a subsequent partial surface reflection on the back side of the star, the contribution to the amplitude spectrum can be expressed as:

\begin{equation}
\sum_{\omega,\ell} \left( V_{\omega,\ell} A_{\omega, \ell} + A'_{\omega, \ell} e^{i\delta} 
\left[V'_{\omega,\ell} e^{-i \omega/\Delta\nu_{\omega,\ell} }+
V''_{\omega,\ell} R(\omega) e^{-2i\omega/\Delta\nu_{\omega,\ell}} \right] \right)
\; ,
\label{eq:interf1}
\end{equation}
where $V_{\omega,\ell}$, $V'_{\omega,\ell}$ and $V''_{\omega,\ell}$ are visibility factors, and $\delta$ is a phase
constant that depends on the type of radiation emitted.
The reflection coefficient $R$ is expected to be a monotonically decreasing function of frequency \citep[for further details see][]{1993ASPC...42...15K}. We also assume that, for stochastically excited waves, the  
amplitudes and the visibility factors will be a smooth function of frequency. 
Hence, the amplitude spectrum will be modulated with the periodic functions $\cos(\omega/\Delta\nu_{\omega,\ell})$ and 
$\cos(2\omega/\Delta\nu_{\omega,\ell})$. 
Although in principle different values of $\Delta\nu_{\omega,\ell}$ can be expected in different frequency ranges and 
for different degrees $\ell$, their differences are too weak, and in the present study we fitted the data to just two periodic components,
$\Delta\nu_2=\Delta\nu_{\omega,\ell}$ and $\Delta\nu_1 =\Delta\nu_{\omega,\ell}/2$.
In addition, in some stars one of the signals could be masked by low-visibility factors or low reflection coefficients. 

\begin{figure}[!htb]
\begin{center}
\includegraphics[scale=0.4]{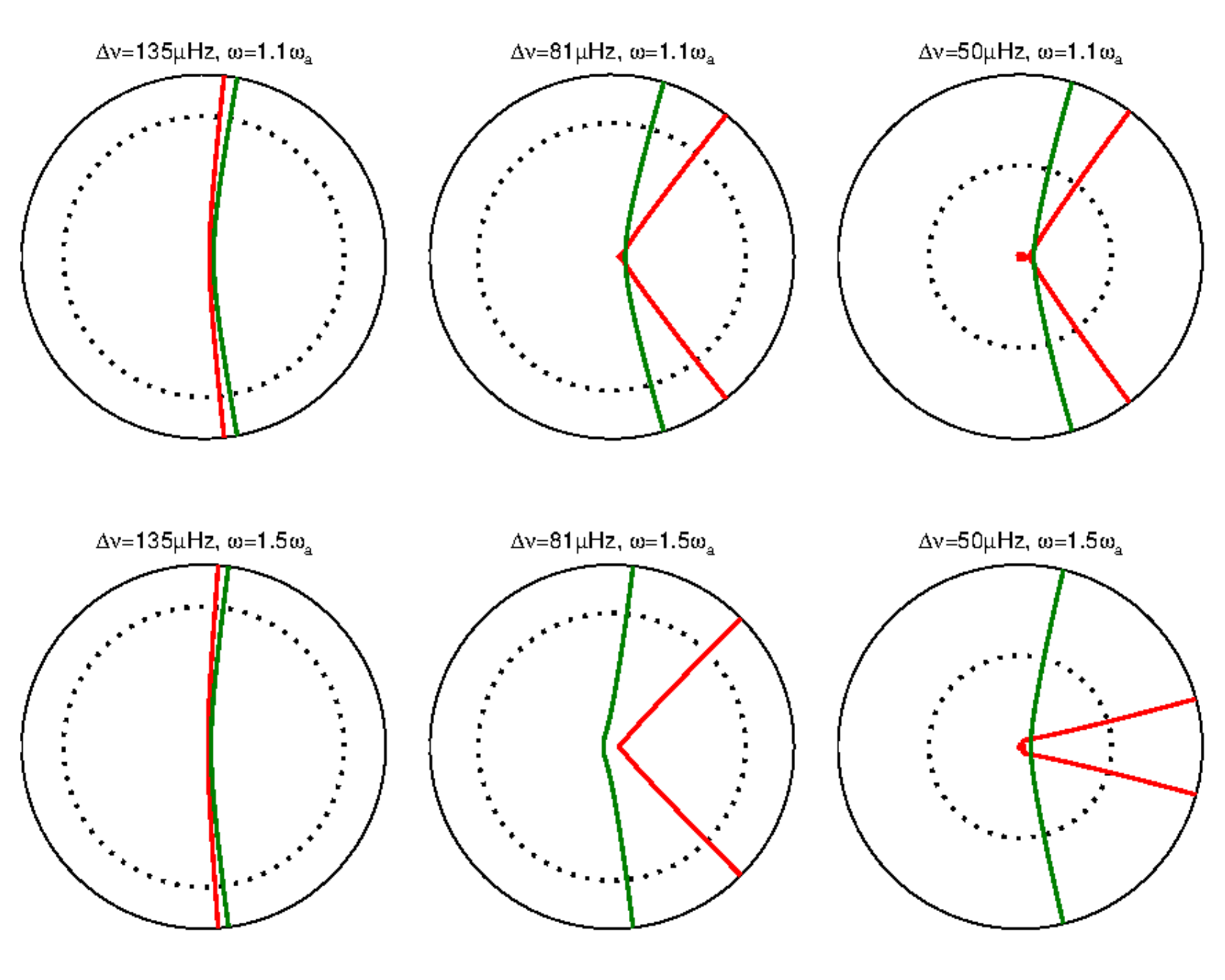}
\caption{ Ray path for waves with angular degrees $\ell=1$ (red) and $\ell=2$ (green) for a $1.1M_{\odot}$ model 
evolution sequence. For clarity, surface reflections have been omitted. 
The upper part shows rays with a frequency a 10\% greater than the acoustic cut-off frequency 
$\omega_{\rm a}$ while the bottom panels corresponds to rays with a frequency 50\% greater than $\omega_{\rm a}$. 
The large separation indicated is the integral $(2\int_0^R 1/c)^{-1}$ and can be used to identified the models in 
Fig.~\ref{fig:angle}. The leftmost panels correspond to a model near the 
ZAMS, the middle panels are for a model near the TAMS, and the rightmost panels are for a model in the RGB. 
The dotted circles indicate the base of the convection zone.}
\label{fig_Sun_rays}
\end{center}
\end{figure}

To illustrate the problem we have considered a $1.1M_{\odot}$ evolution sequence and plotted in Fig.\ref{fig_Sun_rays} 
rays for two typical frequencies in the pseudo-mode
range and three evolutionary stages. Top panels are for frequencies $\omega=1.1\omega_a$ while
bottom panels are for $\omega=1.5\omega_a$. On the other hand the star evolves from left (ZAMS) to right (RGB).
For clarity, surface reflections have been omitted.

For the model close to the ZAMS (leftmost panels) the inward rays emerge at the surface making angles between
$160^{\circ}$ and $180^{\circ}$ with the outward component. In this case, the outward and inward components of a given
wave can be simultaneously visible only if they lie too close to the limb. Hence, their signal 
corresponding to $\Delta\nu_2 \simeq \Delta\nu $ will be highly attenuated in the power spectrum. 
On the other hand, after one surface reflection the angle 
between the outward and inward components lies in the range $0^{\circ}$ to $35^{\circ}$. This interference would give the  
$\Delta\nu_1\simeq \Delta\nu/2$ separation 
and, hence, had it the same intrinsic amplitude as the former, the $\Delta \nu_1$ separation would be easier to observe. 
But in this case the inward rays are only partially reflected on the far side (e.g.\ \cite{1993ASPC...42...15K}) and hence 
the attenuation factor, corresponding to $R(\omega)$ in Eq.~\ref{eq:interf1}, is high. Since the actual observations 
show that in the Sun the dominant separation in the HIP 
pattern is $\Delta\nu_1$ \citep{GarPal1998} we can use this case as a qualitative reference between the two competing factors. 
For model S, we obtain $\Delta\nu_1$ = 72 $\mu$Hz with our fit, for which a frequency range $\sim$[5200,6600] $\mu$Hz was used.
This result is in agreement with the observational value of 70.46  $\pm$ 2 $\mu$Hz found by \citet{GarPal1998}. As mentioned in section 4.1, we re-analysed the solar data (GOLF and VIRGO/SPM) and obtained $\Delta\nu_1$$\sim$ 70 $\mu$Hz (see Table~\ref{tab:dnu}), in agreement with previous values given in the literature. At higher frequencies we re-fitted the data with two sinusoidal components and found  $\Delta\nu_2$$\sim$140$\mu$Hz (Table~\ref{tab:dnu}). This value is in agreement with the theoretical prediction already described by Garcia et al.\ (1998) if we assume that the photosphere is the only source of partial wave reflection. However, in this case the amplitude is much smaller because only waves close to the limb contribute to this interference pattern.

As the stars evolve, the angle between the outward ray and the 
first surface appearance of the inward refracted ray becomes smaller for the $\ell=1$ pseudo-modes. 
This is clearly apparent in Fig.~\ref{fig_Sun_rays}.
Hence, the $\Delta \nu_2\simeq \Delta\nu$ interference pattern becomes easier to observe (higher observed $A_2$ amplitudes). 
This phenomenon is due to the transition between acoustic rays smoothly bending through the stellar interior, 
and waves where the density scale height in the stellar inner structure becomes close to their wavelength 
and the ray behaviour at the inner turning point is closer to a two-layer reflection.
Finally, in the most evolved model in Fig.~\ref{fig_Sun_rays} (upper and bottom right panels) the $\ell=1$ waves have a gravity 
character close to the centre that bends the ray into a loop.

In order to reproduce the observed power spectrum schematically we used Eq.~\ref{eq:interf1}, and considered 
ray traces for a continuum spectrum of waves with 
frequencies between the acoustic cut-off frequency at the surface and about a $1.5$ times that value.
This frequency interval approximately spans the observed pseudo-mode 
frequency range. Since we are interesting only in reproducing the periodicities of the pseudo-mode spectra, constant amplitudes 
were considered, and visibility factors either constant or proportional to $\cos \Theta/2$,
where $\Theta$, is the angular distance between the outward and inward waves. 
We determine the periodic signal in the spectrum by a non-linear fit to the above equation.

\begin{figure}[!htb]
\begin{center}
\includegraphics[width=0.48\textwidth]{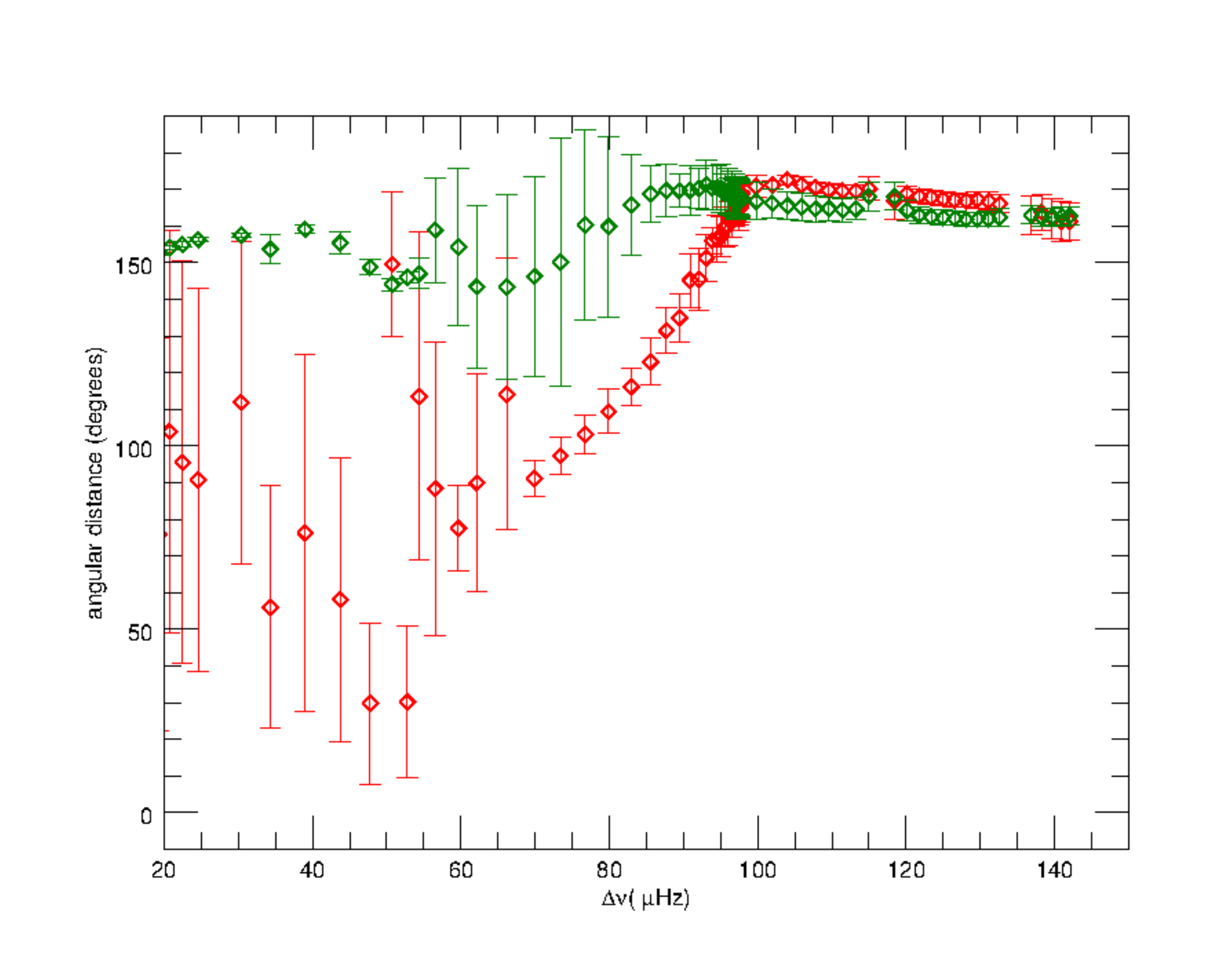}
\caption{Angular distance between inward and outward waves against $\Delta \nu$ for a 
$1.1M_{\odot}$ evolution sequence. Frequencies 
between the  acoustic cut-off frequency at the surface and about  $50\%$  greater than 
$\omega_{\rm a}$ are considered. For every model in the sequence,
frequency-averaged values of the angular distance and $1\sigma$ dispersions are calculated 
for $\ell=1$ (red points) and $\ell=2$ (green points) waves separately. 
An angle of $180^{\circ}$ means no inner reflection.}
\label{fig:angle}
\end{center}
\end{figure}

Fig.~\ref{fig:angle} shows the angular distance between the inward and outward components before any 
surface reflection against the large separation $\Delta\nu$ for models in a $1.1M_{\sun}$, $Y=0.28$ evolution sequence with 
the code and physics indicated in section \ref{sec:models}.
Here, mean values of the angular distance are computed by averaging those of the ray traces in the frequency range 
indicated above. Pseudo-modes with $\ell=1$ and 2 are averaged separately.
Since no surface reflection is considered, these waves form the $\Delta \nu_2$ pattern.

Looking at the behaviour of the $\ell=1$ pseudo-modes in Fig.\ref{fig:angle} (red points) we may conclude that 
for $M=1.1M_{\odot}$ stars that evolve to
a $\Delta \nu$ below some given threshold between $90\,\mu$Hz and $60\,\mu$Hz 
(corresponding to evolved main sequence stars
as the one shown in the middle panel of Fig.~\ref{fig_Sun_rays}), the $\Delta \nu_2$ pattern should become 
visible. Indeed, a smaller angular distance means more disc-centred interferences favouring a higher $A_2$. 
The critical $\Delta\nu$ slightly changes with mass, being lower for higher masses. 
In any case all the stars in Table~\ref{tab:dnu}, except the Sun, are evolved to a point where the pattern with $\Delta\nu_2$  
could be easily observed, as is in fact the case. 

Regarding the $\ell=2$ pseudo-modes, it can be seen in Fig.~\ref{fig:angle} that the angle distance always remains 
above some $150^{\circ}$ and hence this degree contributes only to the $\Delta\nu_1$ period in the pseudo-mode
spectrum. The higher dispersion in the angular distance for models with a large 
separation between $\Delta\nu=60$ and $90\mu$Hz (evolved main sequence stars and subgigants), clearly visible in 
Fig.~\ref{fig:angle}, deserves a comment.
According to the dispersion relation used in this study, for models evolved near the TAMS
the characteristic frequencies of the $\ell=2$ waves become complex in a small radius interval above the core, thus
increasing the resonant cavity of the pseudo-modes within a limited frequency range. An example of the ray path of these kinds
of waves is shown at the bottom middle panel of Fig. \ref{fig_Sun_rays} (green line).
Here, the ray trace close to the centre is not of the acoustic type and of the emerging ray is consequently scattered. Although
these results may be questionable in terms of the validity of the approximation used here, they do not have any observable consequences.

For the three most evolved stars in our dataset the interference pattern coming from waves reflected back at 
the surface with period $\Delta\nu_1$ is not observed.
Because, as noted before, for evolved stars the signal from the $\ell=1$ pseudo-modes do not suffer any attenuation 
from surface reflections, it is possible that this explains why for stars evolved to the RGB the $\Delta\nu_1$ signal becomes 
completely masked. 
However, it is interesting to note the another, possible superimposed, cause for this circumstance.
In principle, one might expect that radial waves propagates all the way
from the surface to the centre and hence contributes to the signal only once reflected back, thence with a period
$\Delta\nu_1$. However 
there is a point in the evolution where the cut-off frequency in the core rises above the 
typical pseudo-mode frequencies, in which case the ray theory introduces an inner reflection.
Although a plano--parallel approximation is questionable in terms 
of the wavelength--radius relation when we are too close to the centre, this approximation allows us to estimate the transmission coefficient $T$, as for the 
one dimensional problem.\footnote{For radial oscillations 
the full adiabatic equations are of second order and hence a WKB analysis can be done 
without the approximations assumed in the dispersion relation Eq. (\ref{edisp}). The qualitative results given in this 
paragraph rely solely on the assumption of a one-dimensional problem.}
{If the equation $T=1/(1+e^{2K})$ is used with $K=2\int_{0}^{r_1} \sqrt{(\omega_c^2 - \omega^2)/c^2}$, $r_1$ being 
the point where $k_r=0$, 
it happens that, for red giants stars, radial waves in the observed frequency range of
the pseudo-mode spectrum are mostly reflected at the core edge.
Thus at this stage radial pseudo-modes also contribute to the $\Delta\nu_2$ signal and hence 
the $\Delta\nu_1$ period would hardly be observed.

Let us now discuss the relation between the large separation obtained from the eigenmodes, $\dnumodes$ and those corresponding
to the pseudo-modes, $\Delta\nu_1$ and $\Delta\nu_2$.
First, we have verified that for waves with $\ell=0$ the phase travel time is always close to the asymptotic acoustic value, 
$\int kdl \simeq \int_{0}^{R} 1/c\, d r$. Hence, when these acoustic waves contribute to the HIPs after a surface reflection 
they give $\Delta\nu_1\simeq \dnumodes/2$. In addition, most of the $\ell=2$ and, in some stars, $\ell=1$ travelling waves 
have similar acoustic characteristics, which also contributed to the same interference period. 
This is in agreement with Table~\ref{tab:nucut_numax}, where both frequency separations match within the errors.

The case of $\Delta\nu_2$ is different.
The grey points in Fig.~\ref{fig:phase} are the periods of the interference pattern computed according to
Eq.~\ref{eq:interf1} for travelling waves and no surface reflection. A wide range of masses and evolutionary
stages from main sequence to the base of the red-giant branch were included.
The red points are the observed $\Delta\nu_2$ values  while blue points are $2\Delta\nu_1$ when observed, including 
the Sun. The black line corresponds to $\dnumodes=\Delta\nu_2$. 
From Fig.~\ref{fig:phase} we can see that for $\dnumodes = 60$--$80\,\mu$Hz there are models with 
$\Delta\nu_2 < \dnumodes$. In fact they correspond to evolved stars up to the end of the main sequence or the subgiant phase,
as it is the case for our sample of stars. 
Although the order of magnitude of the phase delay found for our models is similar to the observed one
for this type of star, it is also apparent from the figure that KIC~1124411 ($\dnumodes \sim 70\,\mu$Hz) 
has a large separation --
too high for this delay to appear. We obtain a mass of $\sim 1M_{\odot}$ either by applying the scaling relations to $\numax$ and $\dnumodes$ by doing an isochrone fit. 
The upper right corner of Fig.\ref{fig_Sun_rays} is representative of this star. Hence we expect
to observe the signal $\Delta\nu_2$ from waves reflected in the interior but our computations do not shown any significant 
delay in the phase of the $\ell=1$ compared to the radial oscillations. 
Further work is need here; in particular, the asymptotic theory that we have used could be inadequate for such details.

For the stars with the lower $\dnumodes$ two of them have $\Delta\nu_2 \simeq \dnumodes$ and one star (KIC~11717120)
has $\Delta\nu_2 < \dnumodes$. With our isochrones fitting, we find that the latter is at the base of the RGB as well as
KIC~7799349, thus we have two different results for stars with very similar parameters. 
With our simple simulation for models at this evolutionary stage we have both signals since,
as noted above, radial oscillations are partially reflected at the core edge and for these stars, 
$\Delta\nu_2 \simeq \dnumodes$. The observed period will be a weighted average, but for a better comparison with the 
observations proper amplitudes need to be computed.

\begin{figure}[!htb]
\begin{center}
\includegraphics[width=0.48\textwidth]{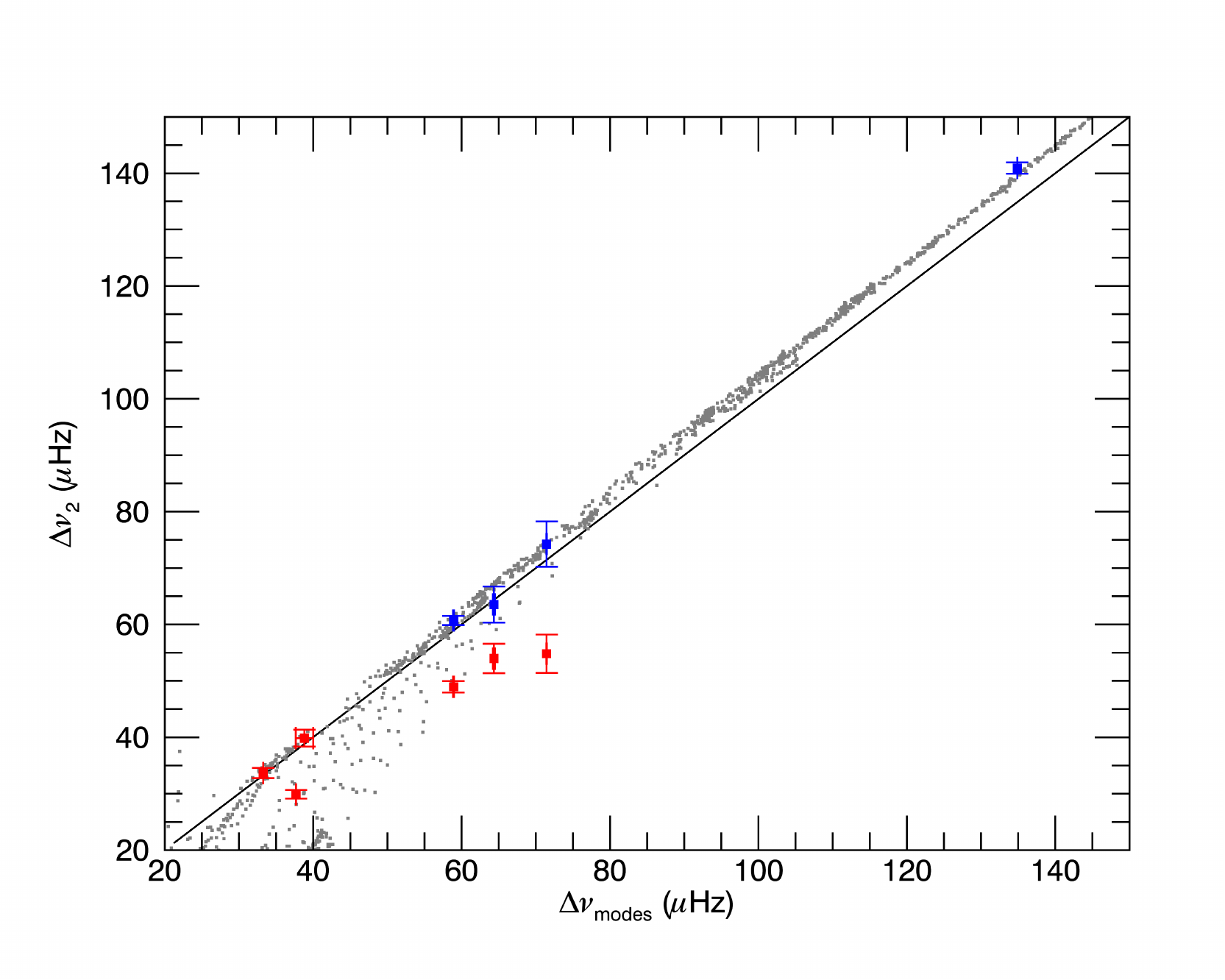}
\caption{Grey points are periods of the pseudo-mode interference pattern for evolution sequences with masses between 
$0.9M_{\odot}$ and $2M_{\odot}$. For the sake of clarity only waves with
no surface reflections are shown. Red points correspond to the observed $\Delta\nu_2$ values and blue points 
correspond to $2\Delta\nu_1$ when observed. The black line corresponds to $ \dnumodes=\Delta\nu_2$}.

\label{fig:phase}
\end{center}
\end{figure}

\section{Conclusions}
In this study the acoustic cut-off frequency and the characteristics of the HIPs have been measured in six star with solar-like pulsations 
 observed by the \emph{Kepler} mission. A comparison with the observed $\numax$ shows a linear trend with all stars 
lying in $\sim$2-$\sigma$. As a result, the values of $\log g$ derived from $\nucut$ agree to within the same accuracy
with those derived from $\numax$ but are substantially different in some cases from the values derived from spectroscopic fits
(see Table~\ref{table:parameters}).
When comparing the stars with the models in the $\nucut$--$\Delta\nu$ plane we found a departure in $\nucut$ from
the expected values. While using theoretical information it is possible to calculate a measurable function 
$\nucut^*=f(\teff, \nucut)$ that scales as $g/\sqrt{\teff}$ with no more than $0.3\%$ deviation 
(for our set of models representative of our star sample and the Sun), we find that the observational $\nucut^*$ does not 
follow this relation so accurately. Rather surprisingly, we found that the frequency of maximum power, $\numax$, follows the 
linear relation within errors. Hence, for the evolved stars considered in the present study 
it must be concluded that $\numax$ gives an even better estimate of $\log g$ than $\nucut$ with the current uncertainties.

Given the observed characteristics of the HIPs, our set of six stars can be divided into two groups. In three stars: KIC~3424541, KIC~7799349, and KIC~11717120, the only visible  pattern is the one due to the interference with 
inner refracted waves in the visible disc of the star and not close to the limb. In the other three stars ( KIC~7940546, KIC~9812850, and KIC~11244118) the pattern due to partial reflection of the inward waves at the back of the star is also detected. 
This different behaviour is related to the wide spacing of the star, which reveals a dependence on the evolutionary stage.

When present, the period $\Delta \nu_1$, which corresponds to the interference of outward waves with their inward counterpart
once reflected on the far side, always agrees with half the large separation, although, like the Sun, their values are
a little higher (this result is also found theoretically, the large separation derived from the pseudo-modes being closer to $\int dr/c$). However, the period $\Delta\nu_2$ corresponding to waves refracted in the inner part 
of the star can be substantially different from the large frequency separation. We interpret this by claiming the presence of
waves of a mixed nature. For these stars, the phase $\int k dl$ is smaller compared to the radial acoustic case. Although a detailed
analysis is beyond the scope of the present study, simple ray theory calculations reveal that such a phenomenon is expected with the same order of magnitude.

The pseudo-mode spectrum reveals information not only from the surface properties of the stars through $\nucut$
but also from the interior, at least in the cases where the period $\Delta\nu_2$ is lower than the large separation 
derived from the eigenmodes. We are aware that the interference periods derived observationally are power-weighted averages 
and for a proper comparison with theoretical expectation some work along these lines should be addressed. This might be accomplished by computing transmission coefficients so that more realistic theoretical simulations of the
interference phenomenon can be done.

\begin{acknowledgements} 
The authors of this paper thank Dr. J. Ballot, Dr. G.R. Davies, and P.L. Pall\'e for useful comments and discussions, as well as  the entire \emph{Kepler} team, without whom these results would not be possible. 
Funding for this Discovery mission is provided by NASA Science Mission Directorate. 
This research was supported in part by the Spanish National Research Plan under project AYA2010-17803.
This research was supported in part by the National Science Foundation under Grant No. NSF PHY05-51164. RAG has received 
funding from the European Community Seventh Framework Program (FP7/2007-2013) under grant agreement no. 269194 (IRSES/ASK), 
from the ANR (Agence Nationale de la Recherche, France) program IDEE (no. ANR-12-BS05-0008) ``Interaction Des \'Etoiles et des 
Exoplan\`etes'', and from the CNES. SM acknowledges the support of the NASA grant NNX12AE17G.

\end{acknowledgements}

\appendix
\section{Figures of the other stars}

\begin{figure}[!htb]
\begin{center}
\includegraphics[scale=0.4]{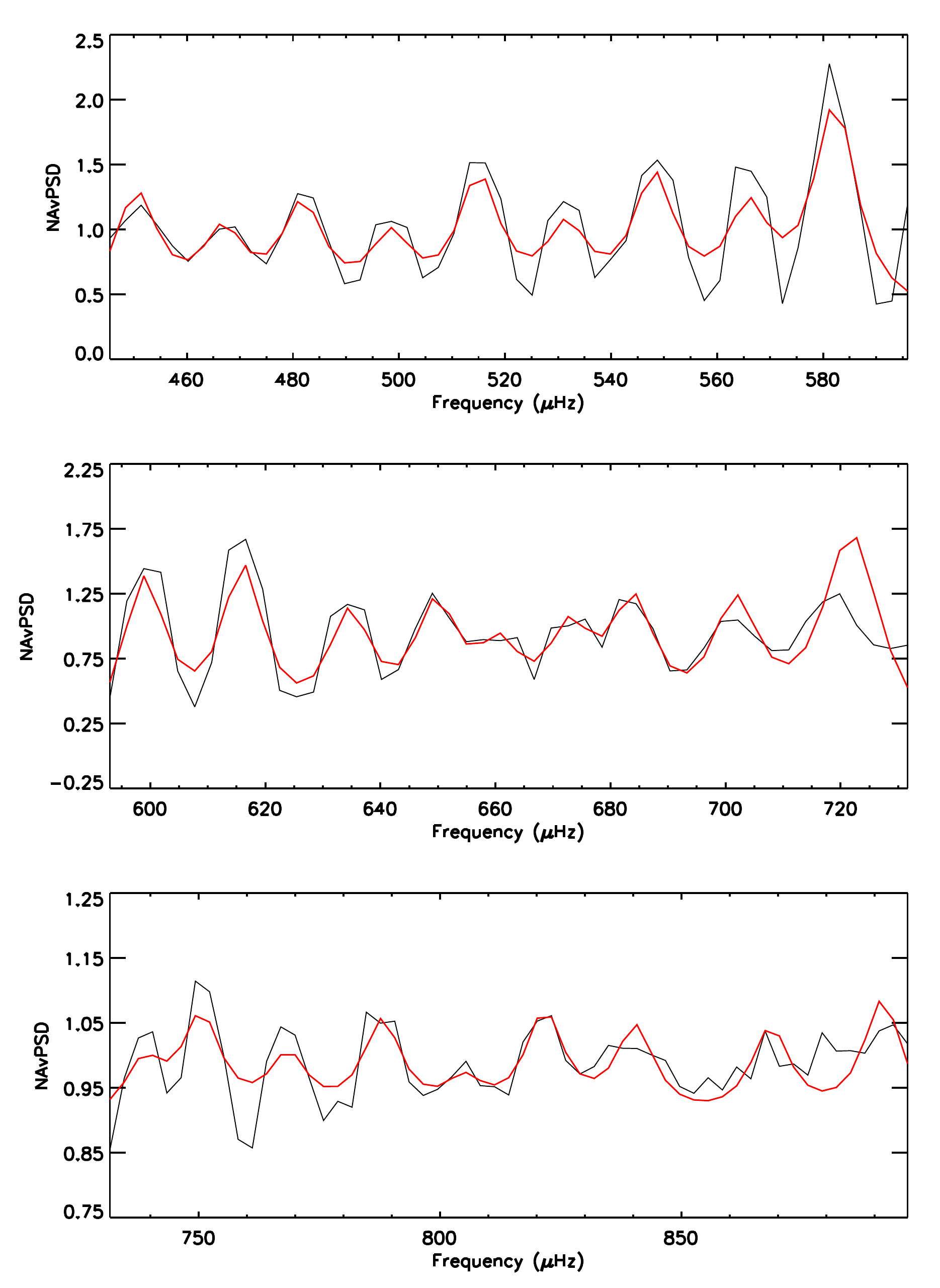}
\caption{Fitted (red line) NAvPSD  of KIC~7799349 by groups of nine peaks in the low-frequency range (p modes).}
\label{fig:numax_nucut1}
\end{center}
\end{figure}

\begin{figure}[!htb]
\begin{center}
\includegraphics[scale=0.3,angle=90]{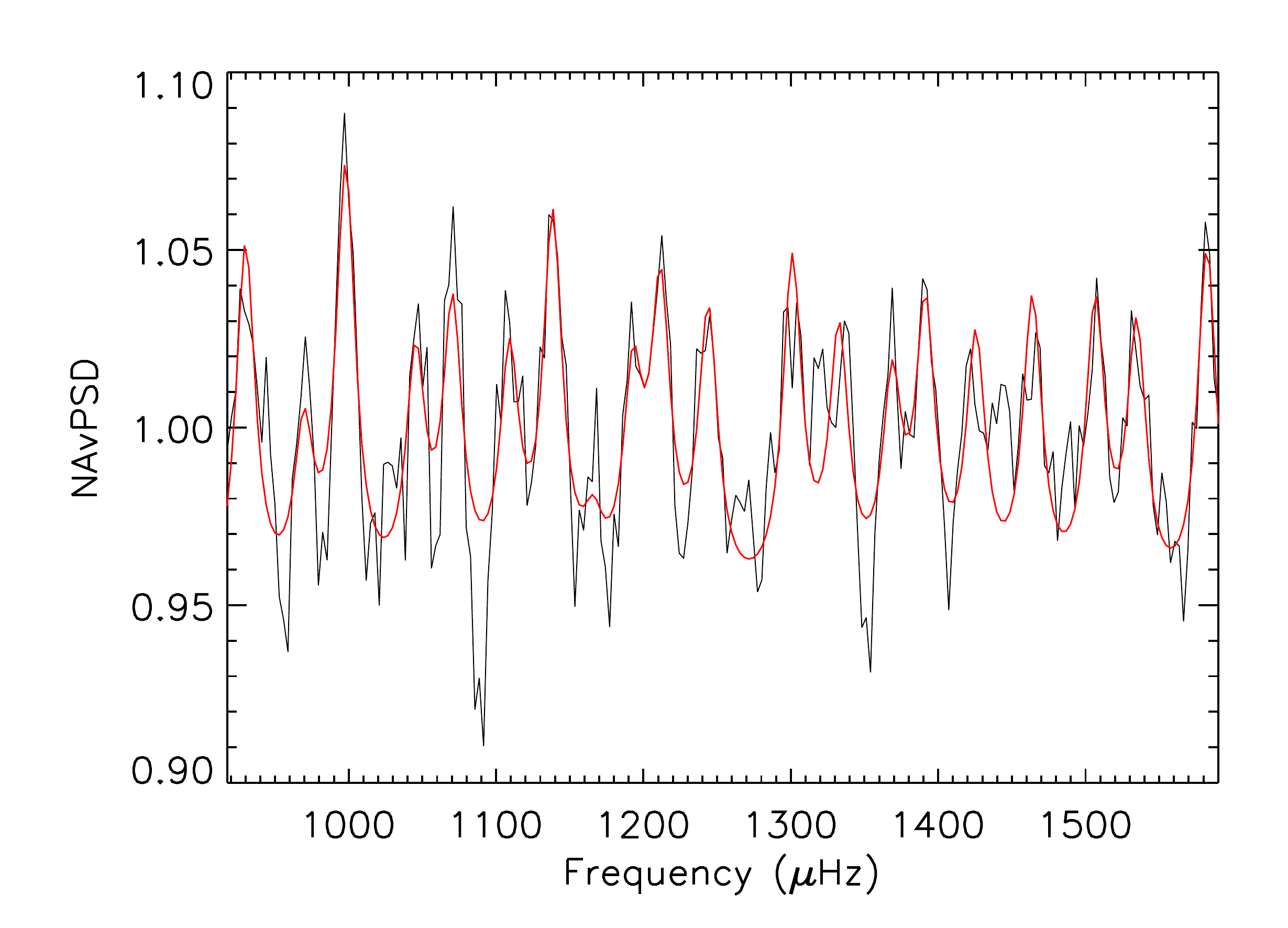}
\caption{Fitted (red line) NAvPSD of KIC~7799349 in the high-frequency range (pseudo modes). }
\label{fig:numax_nucut2}
\end{center}
\end{figure}

\begin{figure}[!htb]
\begin{center}
\includegraphics[scale=0.3,angle=90]{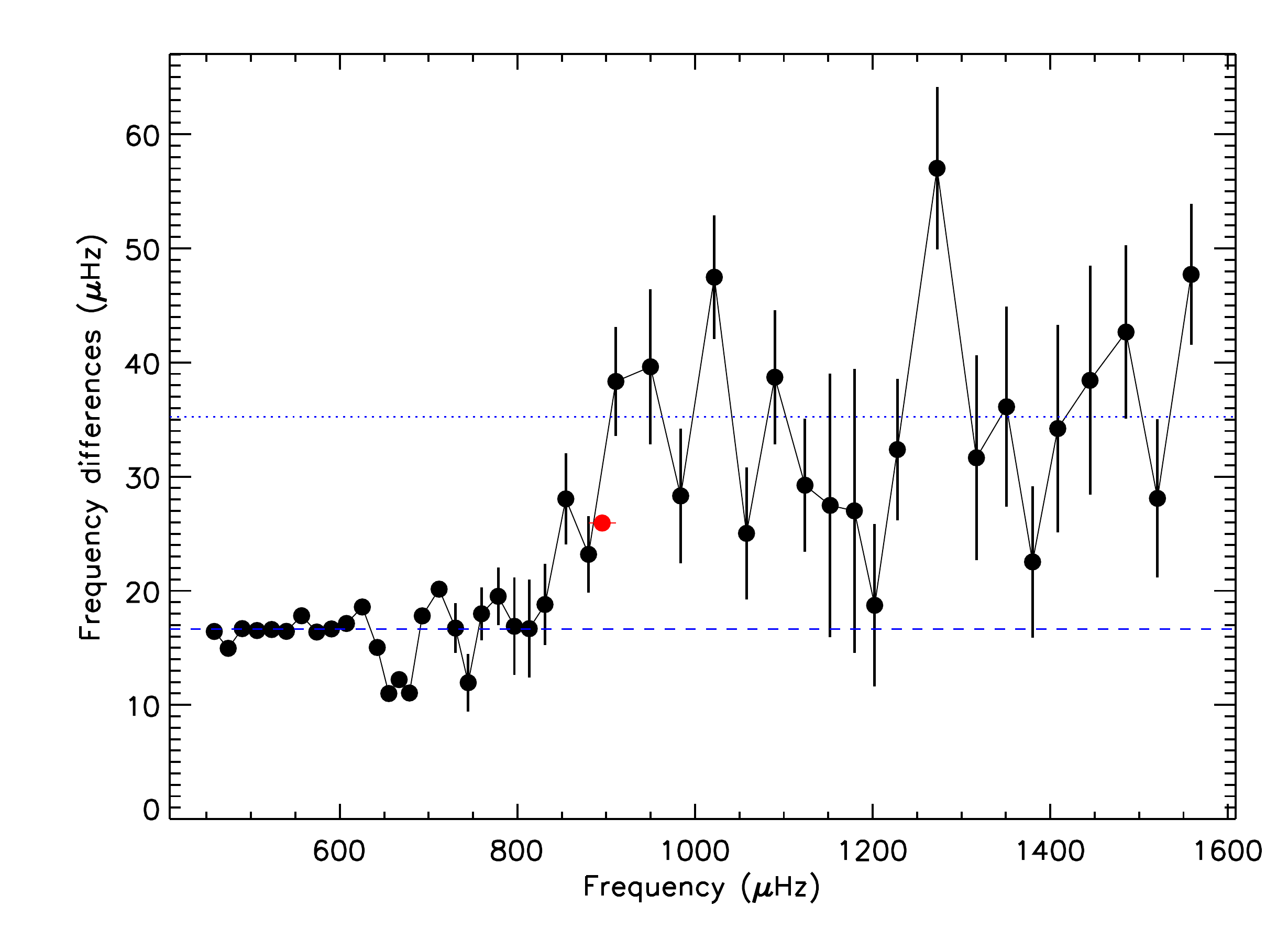}
\caption{Consecutive frequency differences (KIC~7799349); that is, the separations between the fitted peaks for NAvPSD. Two levels appear, one around $\Delta\nu/2$ corresponding to p-modes (blue dashed line with a weighted mean of 16.64 $\pm$ 0.09 $\mu$Hz) and other around $\Delta\nu$ corresponding to pseudo-modes (blue dotted line with a weighted mean of 35.23 $\pm$ 1.52 $\mu$Hz). The red symbol is the estimate of the acoustic cut-off frequency (895.47 $\pm$ 15.38 $\mu$Hz). }
\label{fig:numax_nucut3}
\end{center}
\end{figure}


\begin{figure}[!htb]
\begin{center}
\includegraphics[scale=0.4]{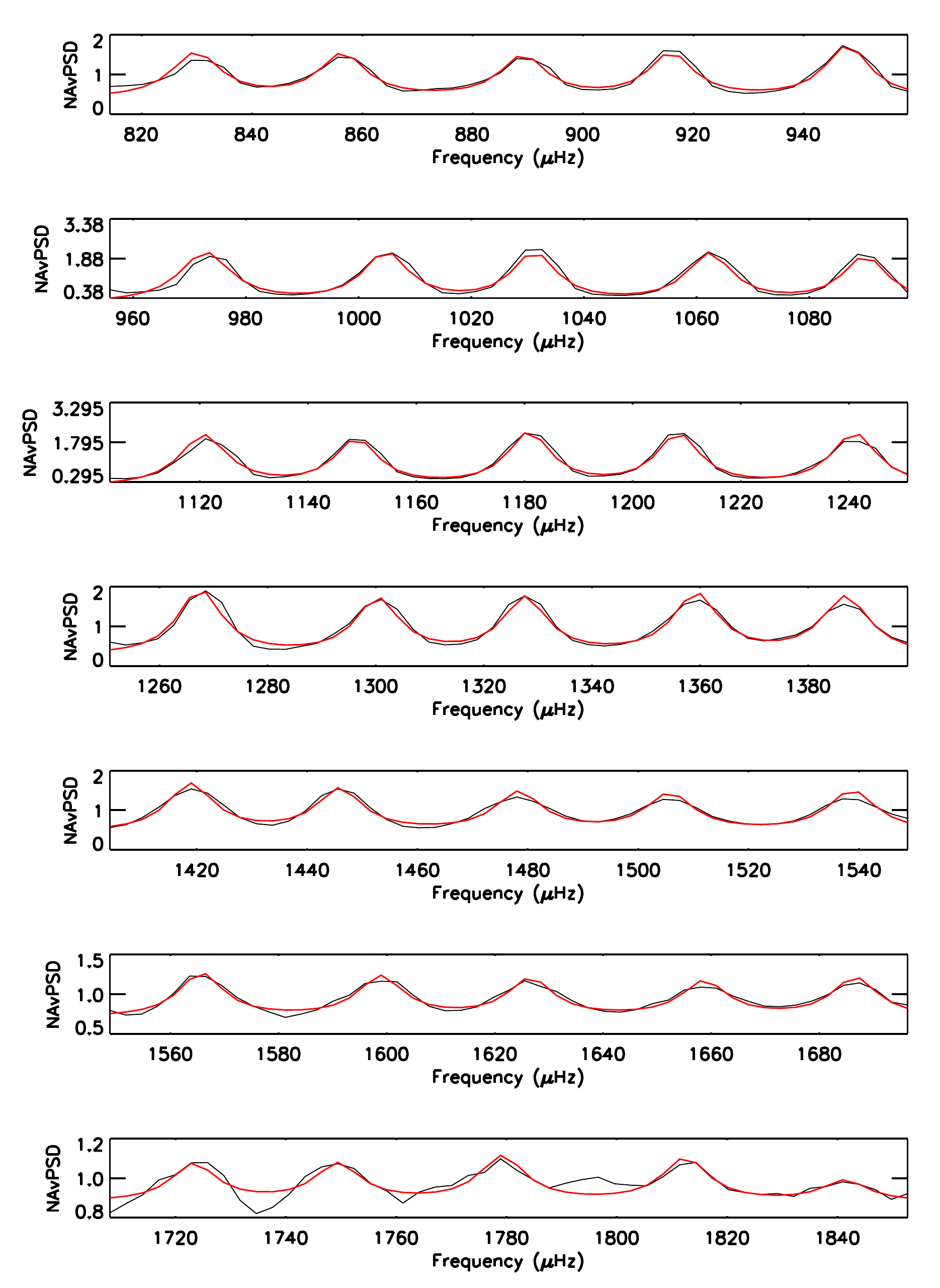}
\caption{Fitted (red line) NavPSD  of KIC~7940546 by groups of five peaks in the low-frequency range (p modes).}
\label{fig:numax_nucut4}
\end{center}
\end{figure}

\begin{figure}[!htb]
\begin{center}
\includegraphics[scale=0.3,angle=90]{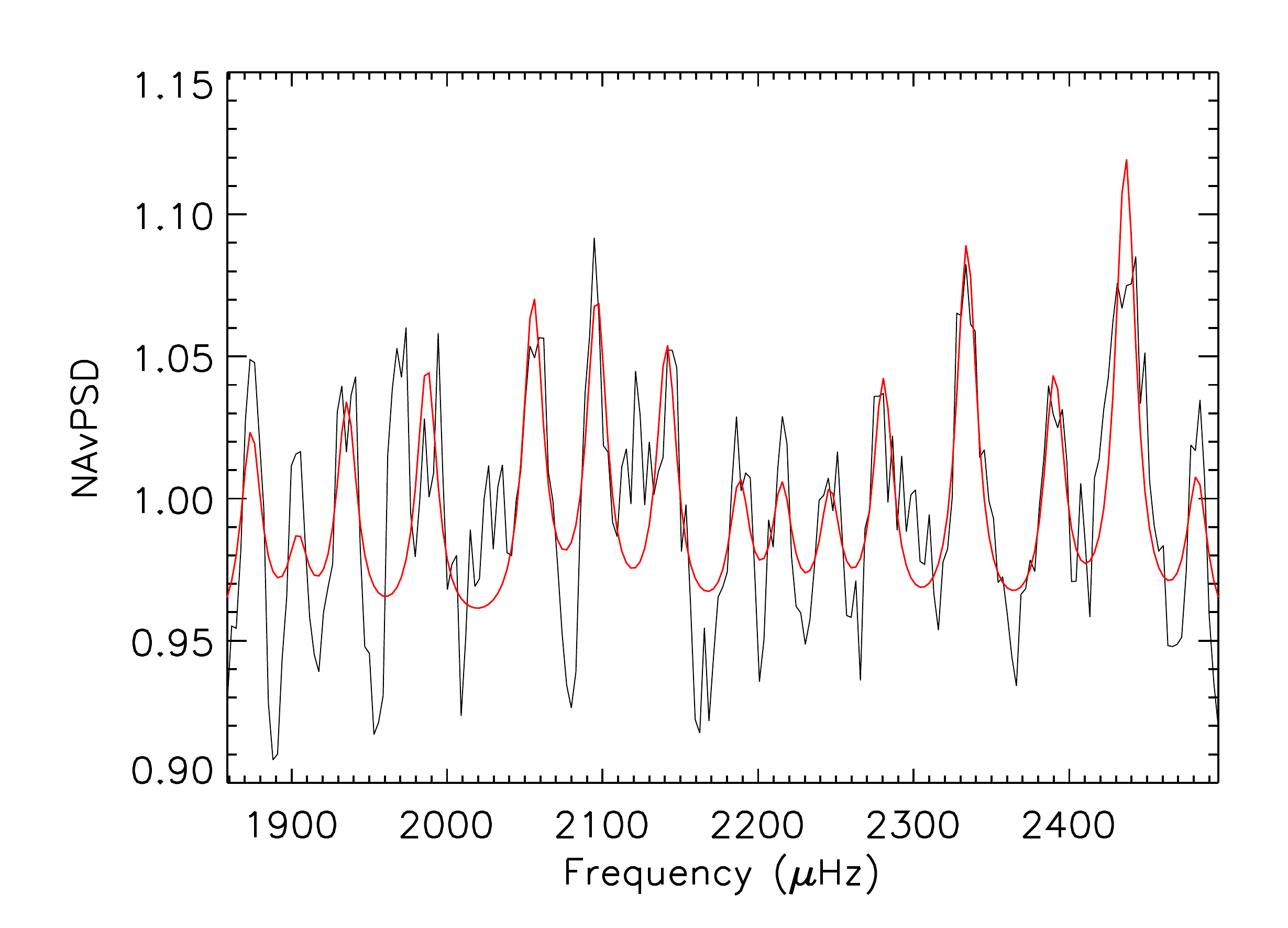}
\caption{Fitted (red line) NAvPSD of KIC~7940546 in the high-frequency range (pseudo-modes). }
\label{fig:numax_nucut5}
\end{center}
\end{figure}

\begin{figure}[!htb]
\begin{center}
\includegraphics[scale=0.3,angle=90]{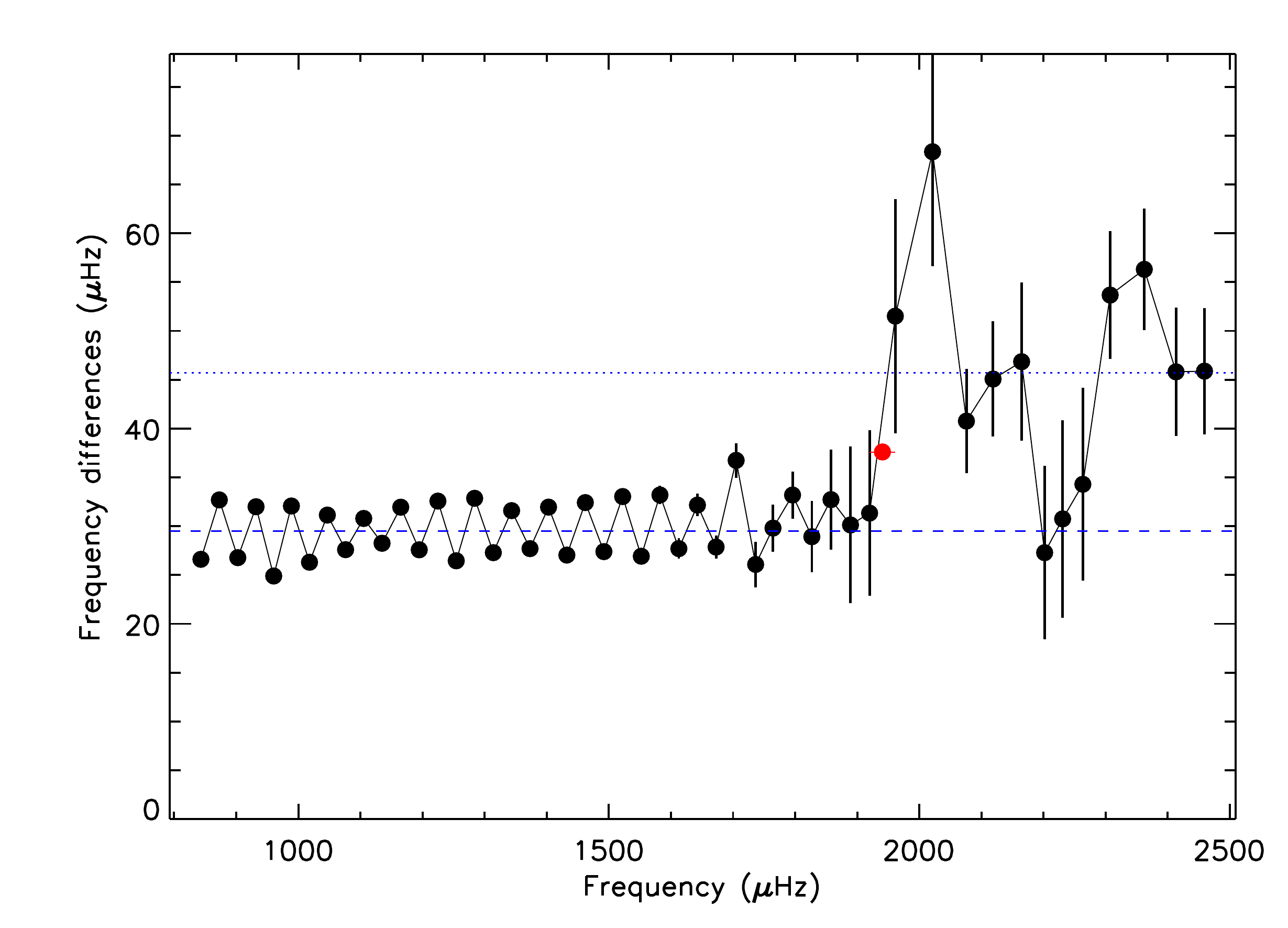}
\caption{Consecutive frequency differences (KIC~7940546), that is the separations between the fitted peaks for NAvPSD. Two levels appear, one around $\Delta\nu/2$ corresponding to p-modes (blue dashed line with a weighted mean of 29.50 $\pm$ 0.07 $\mu$Hz) and other around $\Delta\nu$ corresponding to pseudo-modes (blue dotted line with a weighted mean of 45.71 $\pm$ 2.12 $\mu$Hz). The red symbol is the estimate of the acoustic cut-off frequency (1940.59$\pm$ 20.71 $\mu$Hz). }
\label{fig:numax_nucut6}
\end{center}
\end{figure}


\begin{figure}[!htb]
\begin{center}
\includegraphics[scale=0.4]{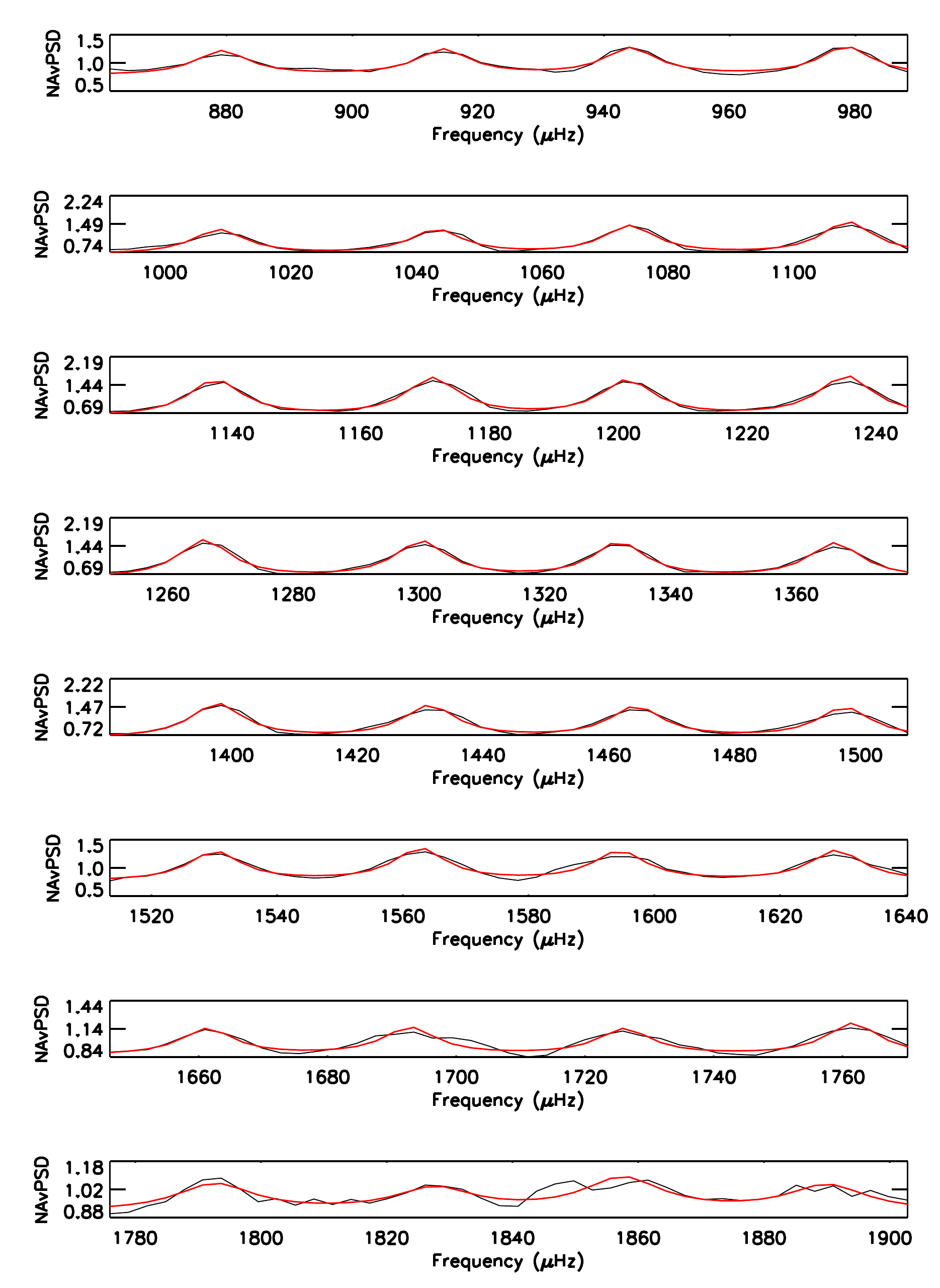}
\caption{Fitted (red line) NAvPSD  of KIC~9812850 by groups of four peaks in the low-frequency range (p modes).}
\label{fig:numax_nucut7}
\end{center}
\end{figure}

\begin{figure}[!htb]
\begin{center}
\includegraphics[scale=0.3,angle=90]{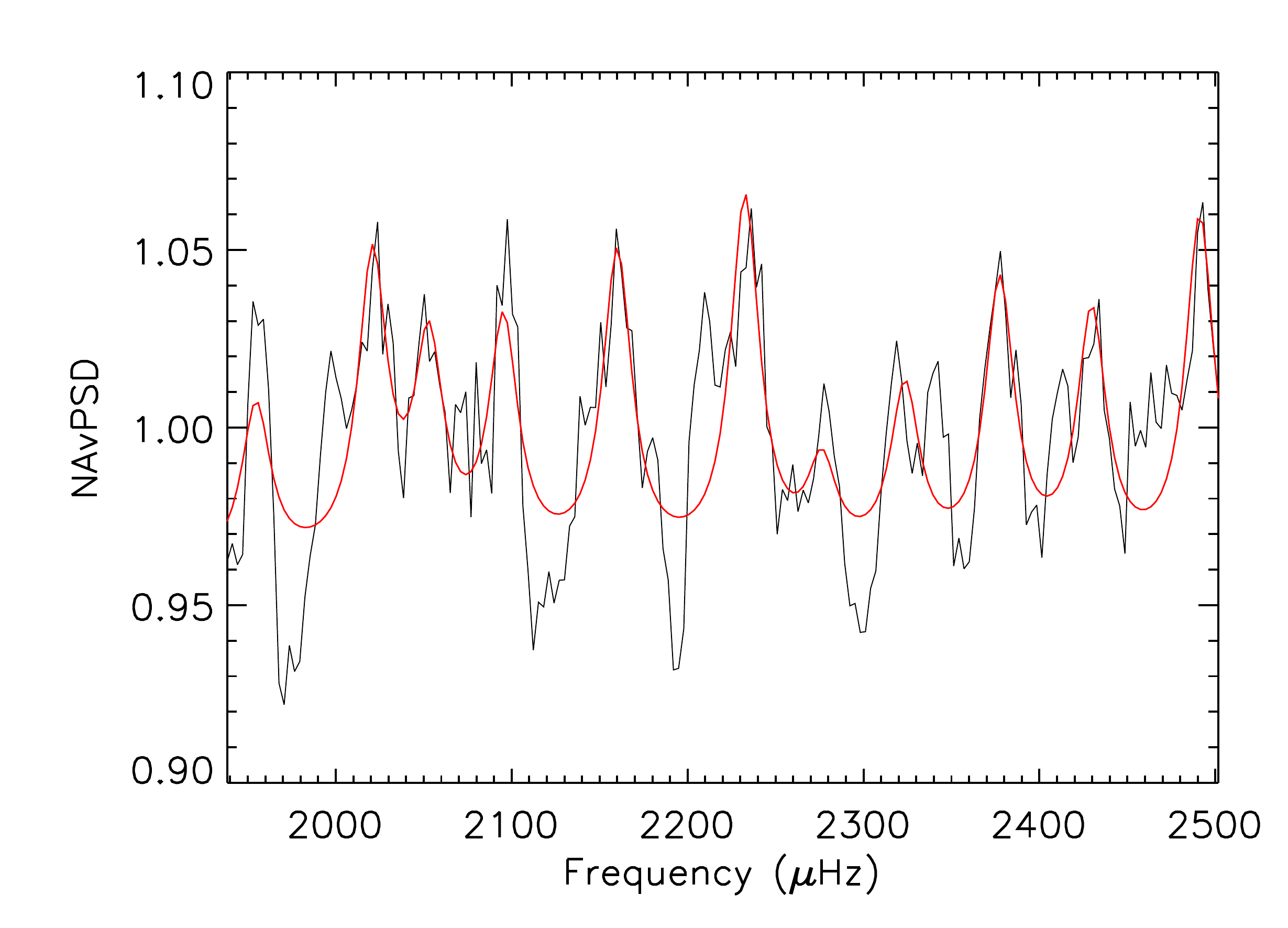}
\caption{Fitted (red line) NAvPSD of KIC~9812850 in the high-frequency range (pseudo-modes). }
\label{fig:numax_nucut8}
\end{center}
\end{figure}

\begin{figure}[!htb]
\begin{center}
\includegraphics[scale=0.3,angle=90]{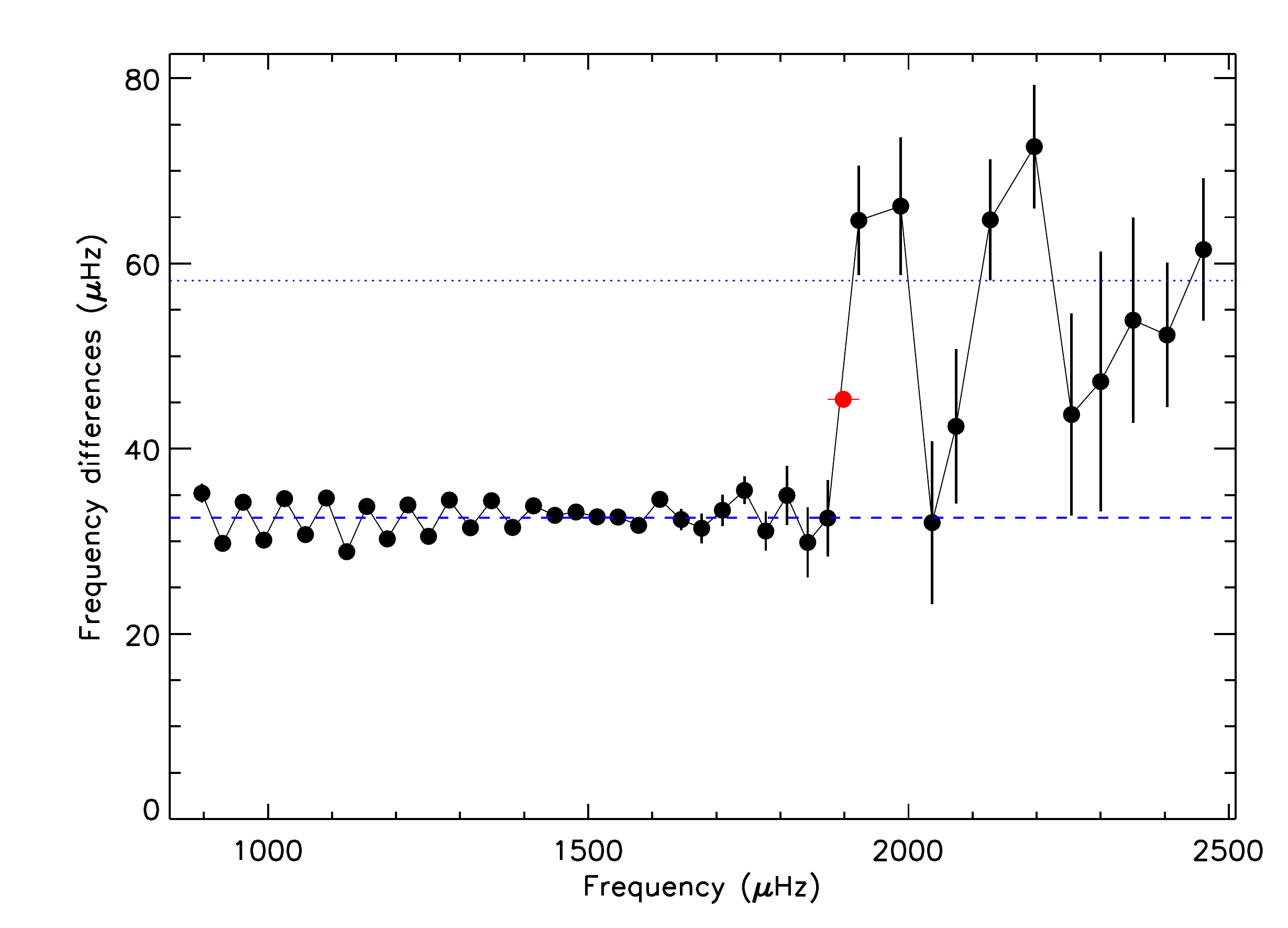}
\caption{Consecutive frequency differences (KIC~9812850); that is, the separations between the fitted peaks for NAvPSD. Two levels appear, one around $\Delta\nu/2$ corresponding to p-modes (blue dashed line with a weighted mean of 32.53 $\pm$ 0.12 $\mu$Hz) and other around $\Delta\nu$ corresponding to pseudo-modes (blue dotted line with a weighted mean of 58.13 $\pm$ 2.39 $\mu$Hz). The red symbol is the estimate of the acoustic cut-off frequency (1898.08 $\pm$ 24.29 $\mu$Hz). }
\label{fig:numax_nucut9}
\end{center}
\end{figure}


\begin{figure}[!htb]
\begin{center}
\includegraphics[scale=0.4]{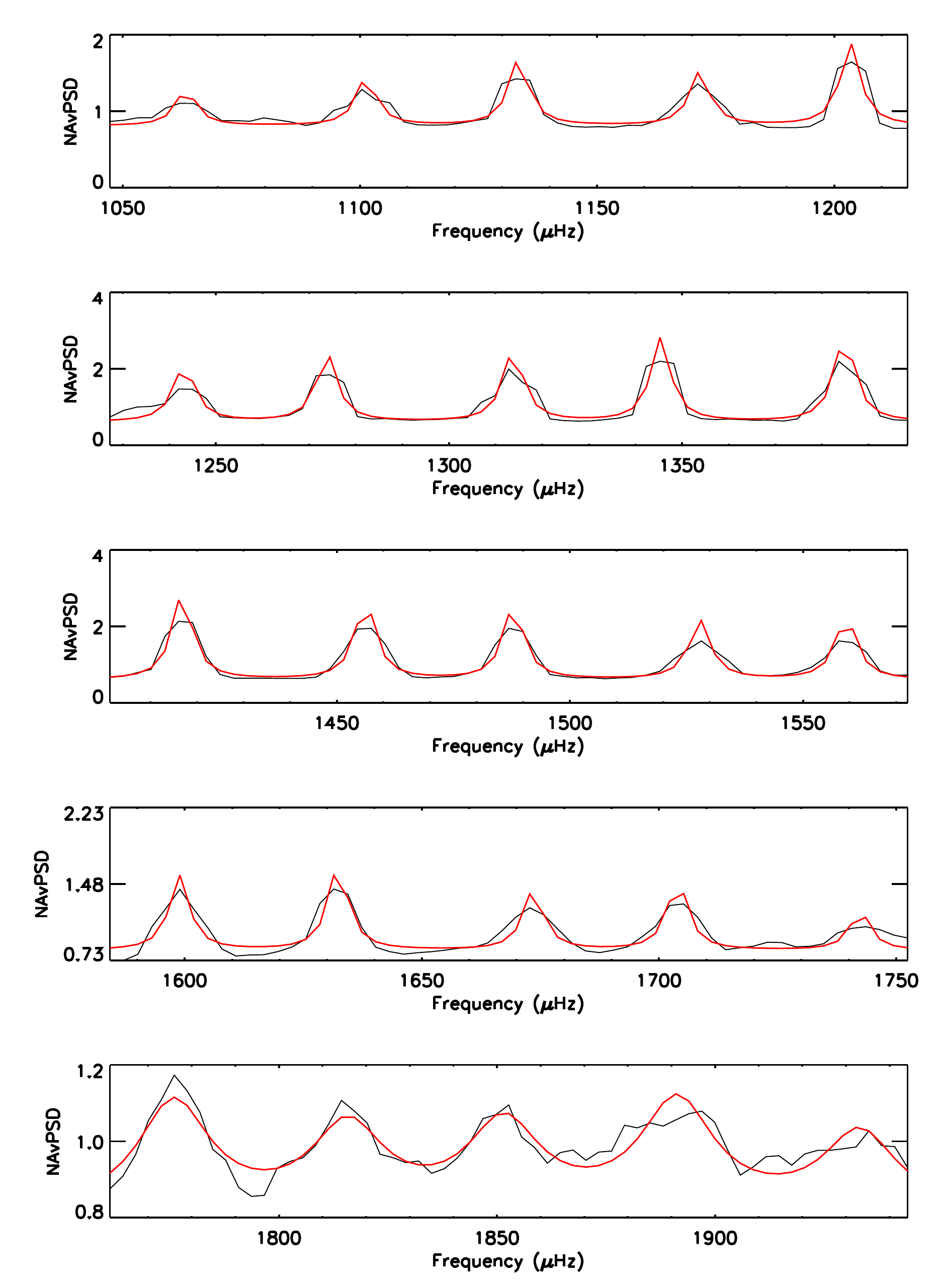}
\caption{Fitted (red line) NAvPSD  of KIC~11244118 by groups of five peaks in the low-frequency range (p modes).}
\label{fig:numax_nucut10}
\end{center}
\end{figure}

\begin{figure}[!htb]
\begin{center}
\includegraphics[scale=0.3,angle=90]{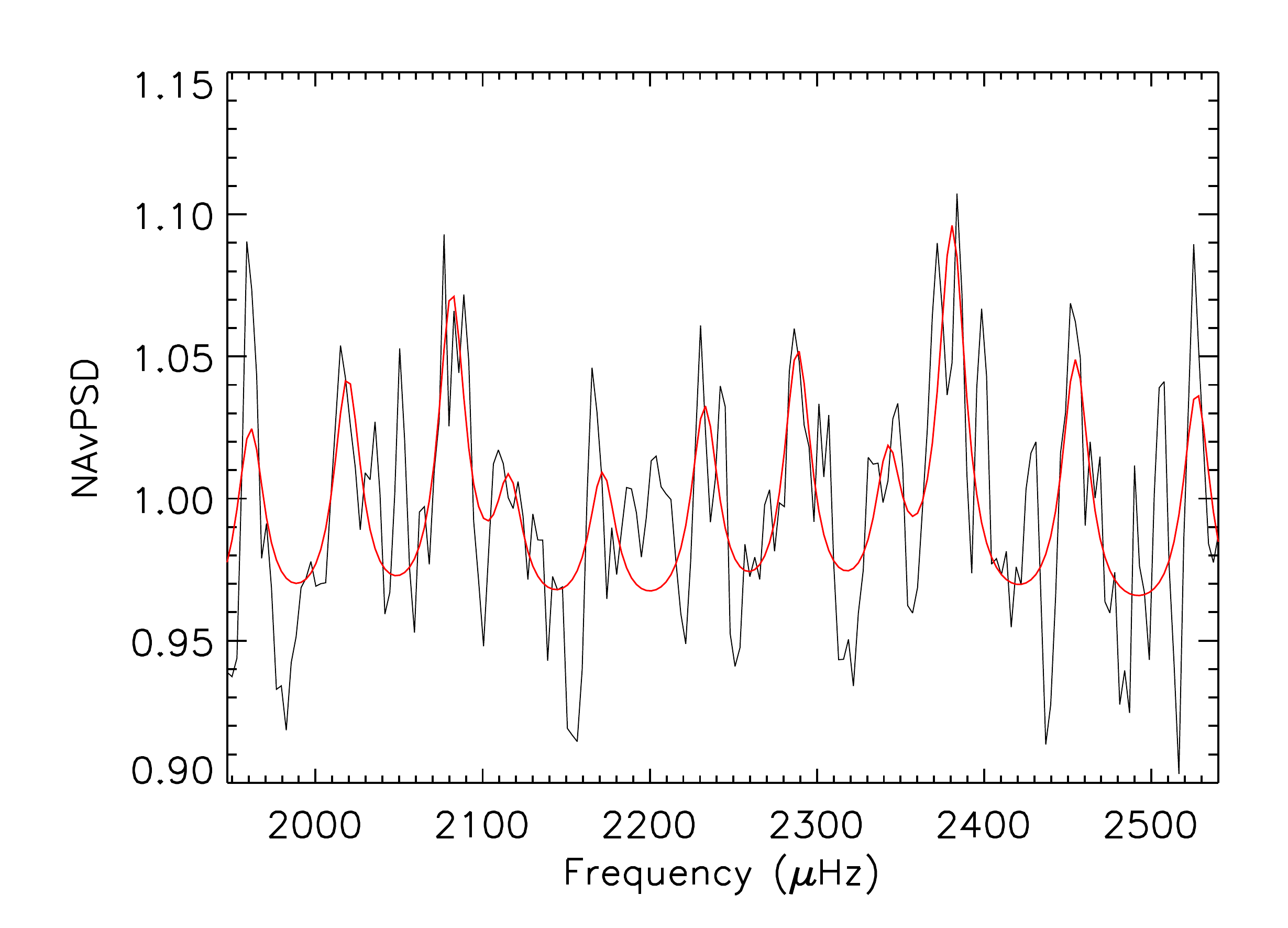}
\caption{Fitted (red line) NAvPSD of KIC~11244118 in the high-frequency range (pseudo modes) }
\label{fig:numax_nucut11}
\end{center}
\end{figure}

\begin{figure}[!htb]
\begin{center}
\includegraphics[scale=0.3,angle=90]{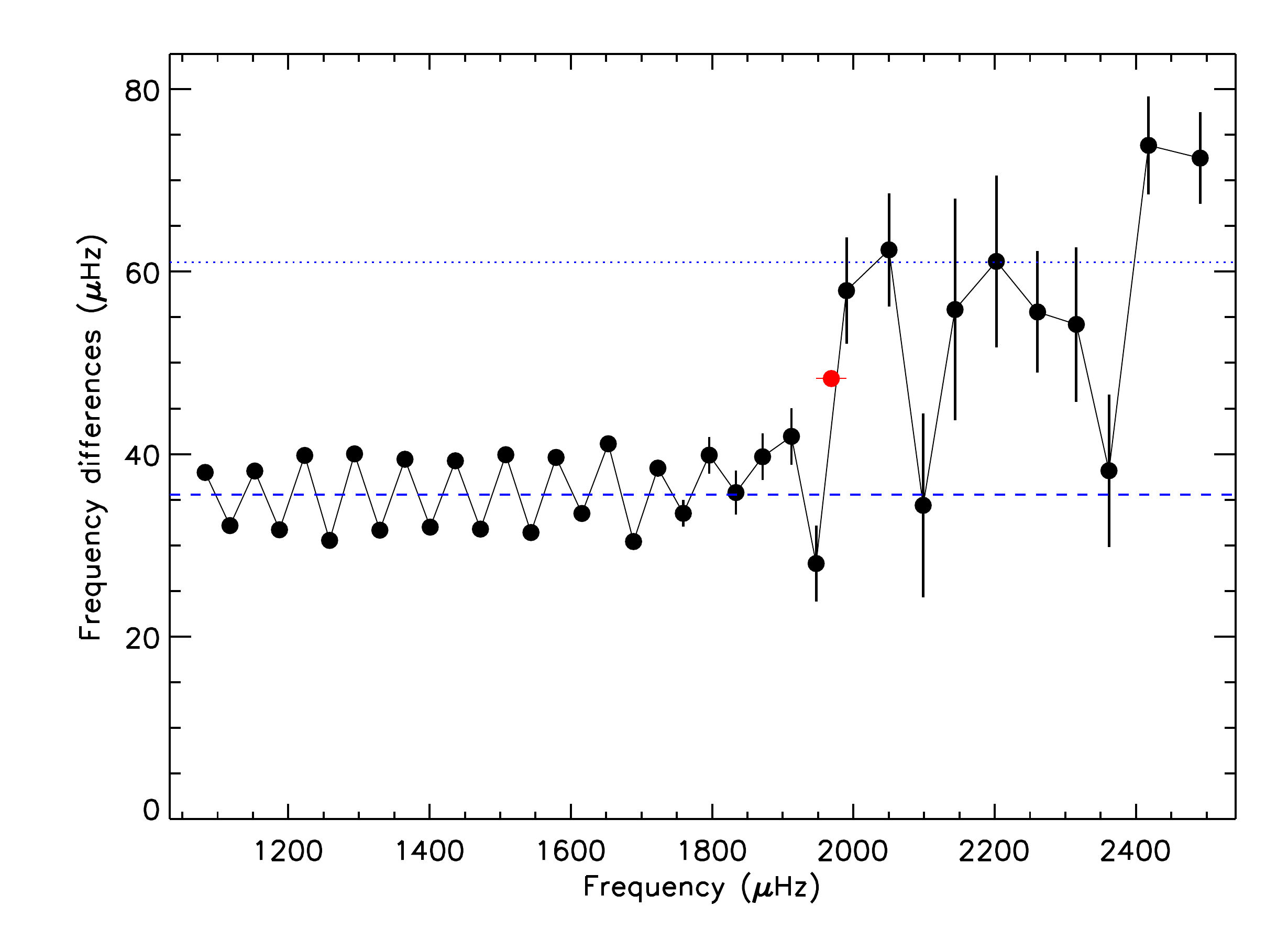}
\caption{Consecutive frequency differences (KIC~11244118); that is, the separations between the fitted peaks for NAvPSD. Two levels appear, one around $\Delta\nu/2$ corresponding to p-modes (blue dashed line with a weighted mean of 35.54 $\pm$ 0.08 $\mu$Hz) and other around $\Delta\nu$ corresponding to pseudo-modes (blue dotted line with a weighted mean of 61.02 $\pm$ 2.19 $\mu$Hz). The red symbol is the estimation of the acoustic cut-off frequency (1968.69 $\pm$ 21.48 $\mu$Hz). }
\label{fig:numax_nucut12}
\end{center}
\end{figure}


\begin{figure}[!htb]
\begin{center}
\includegraphics[scale=0.4]{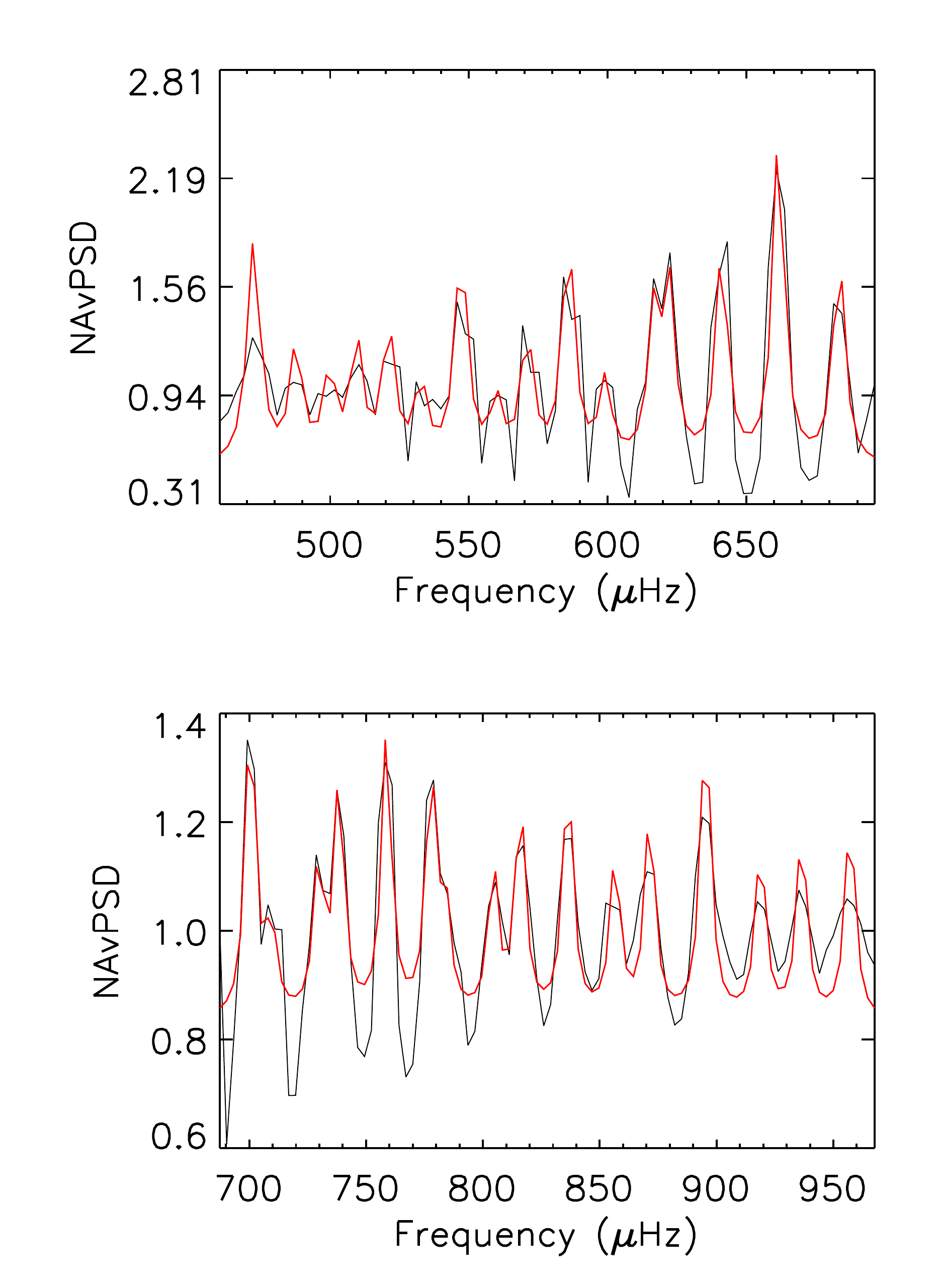}
\caption{Fitted (red line) NAvPSD  of KIC~11717120 by groups of sixteen peaks in the low-frequency range (p-modes).}
\label{fig:numax_nucut13}
\end{center}
\end{figure}

\begin{figure}[!htb]
\begin{center}
\includegraphics[scale=0.3,angle=90]{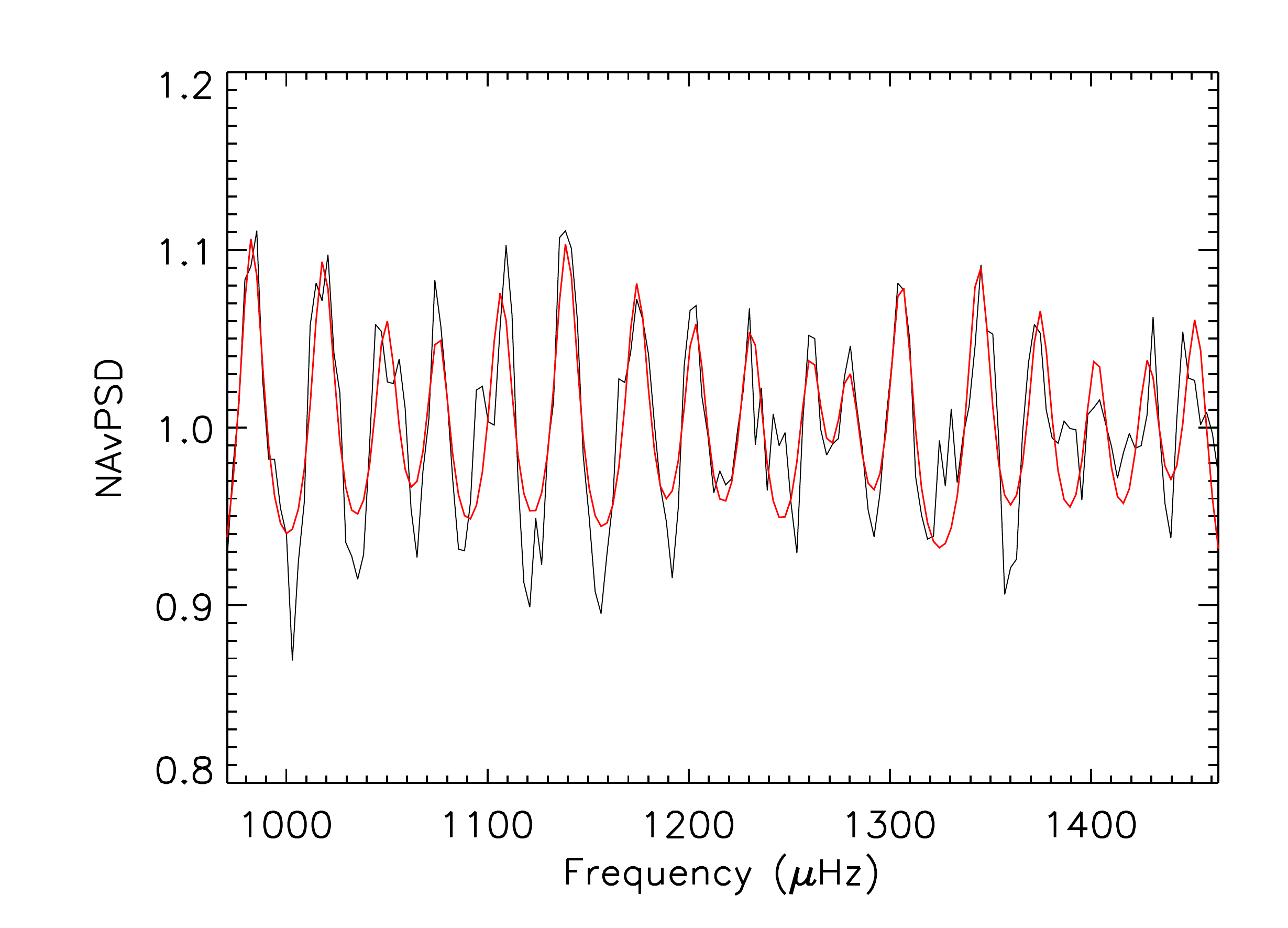}
\caption{Fitted (red line) NAvPSD of KIC~11717120 in the high-frequency range (pseudo-modes). }
\label{fig:numax_nucut14}
\end{center}
\end{figure}

\begin{figure}[!htb]
\begin{center}
\includegraphics[scale=0.3,angle=90]{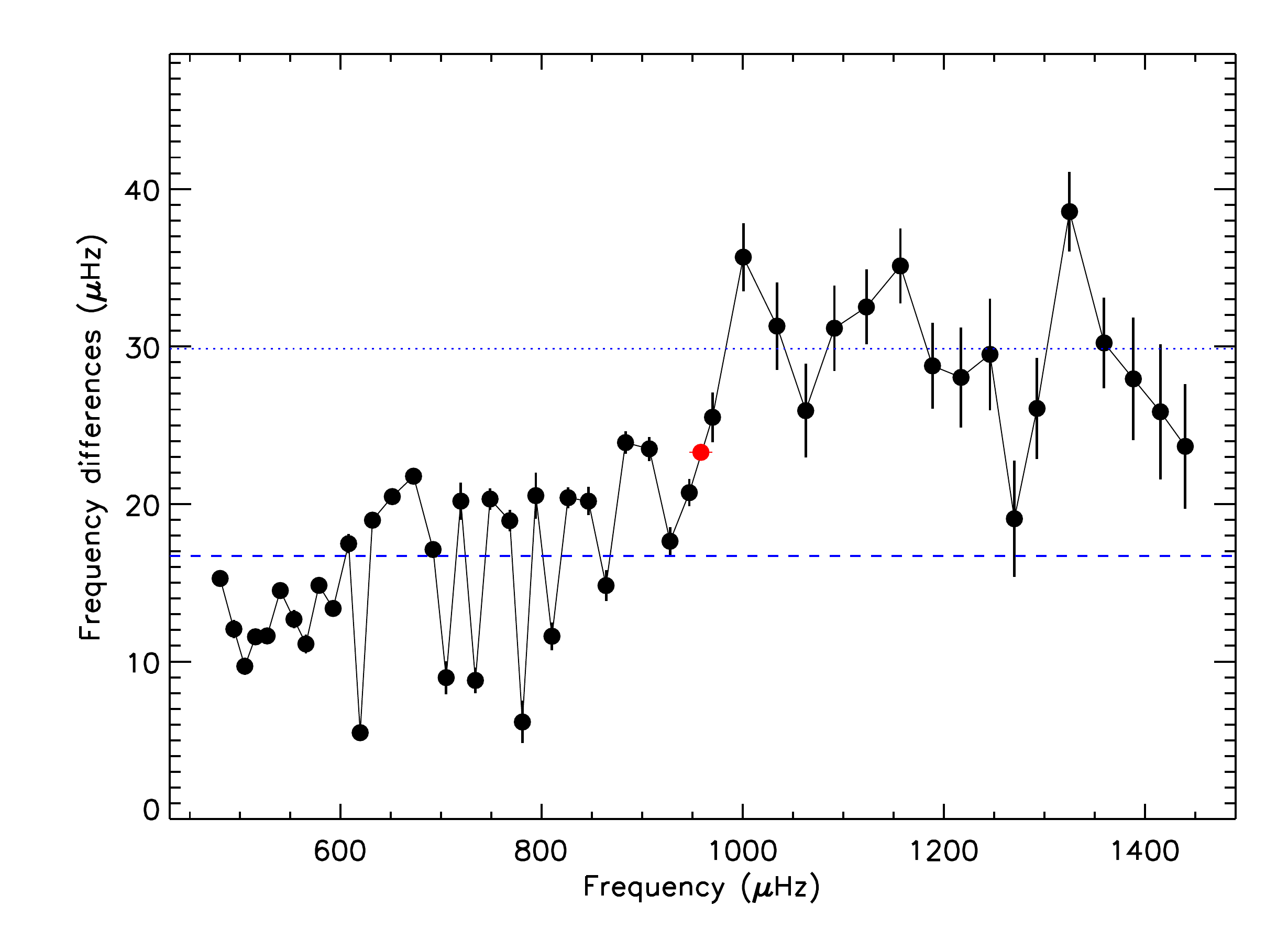}
\caption{Consecutive frequency differences (KIC~11717120); that is, the separations between the fitted peaks for NAvPSD. Two levels appear, one around $\Delta\nu/2$ corresponding to p-modes (blue dashed line with a weighted mean of 16.71 $\pm$ 0.09 $\mu$Hz) and other around $\Delta\nu$ corresponding to pseudo-modes (blue dotted line with a weighted mean of 29.85 $\pm$ 0.66 $\mu$Hz). The red symbol is the estimation of the acoustic cut-off frequency (958.38 $\pm$ 11.56 $\mu$Hz). }
\label{fig:numax_nucut15}
\end{center}
\end{figure}

\bibliographystyle{aa}
\bibliography{./BIBLIO}

\end{document}